Heat treatment – microstructure – hardness relationships of new nickel-rich nickel-titanium-hafnium alloys developed for tribological applications

Sean H. Mills[1], Christopher Dellacorte[2], Ronald D. Noebe[2], Michael J. Mills[3], Aaron P. Stebner[1*], and Behnam Amin-Ahmadi[1†]

[1]Mechanical Engineering Department, Colorado School of Mines, Golden, CO 80401, USA
[2]NASA Glenn Research Center, Materials and Structures Division, Cleveland, OH 44135, USA
[3]Materials Science and Engineering, Ohio State University, Columbus, OH 43212, USA

**ABSTRACT**

The effects of various heat treatments on the microstructure and hardness of new $Ni_{56}Ti_{41}Hf_3$ and $Ni_{56}Ti_{36}Hf_8$ (atomic %) alloys were studied to evaluate the suitability of these materials for tribological applications. A solid-solution strengthening effect due to Hf atoms was observed for the solution annealed (SA) $Ni_{56}Ti_{36}Hf_8$ alloy (716 HV), resulting in a comparable hardness to the $Ni_{56}Ti_{41}Hf_3$ alloy containing 54 vol.% of $Ni_4Ti_3$ precipitates (707 HV). In the $Ni_{56}Ti_{41}Hf_3$ alloy, the maximum hardness (752 HV), achieved after aging at 300°C for 12 h, was attributed to dense, semi-coherent precipitation of the $Ni_4Ti_3$ phase. Unlike the lenticular morphology usually observed within binary NiTi alloys, a "blocky" $Ni_4Ti_3$ morphology formed within $Ni_{56}Ti_{36}Hf_3$ due to a smaller lattice mismatch in the direction normal to the habit plane at the precipitate/matrix interface. The maximum hardness for $Ni_{56}Ti_{36}Hf_8$ (769 HV) was obtained after applying an intermediate aging step (300°C for 12 h) followed by normal aging (550 °C for 4 h). This two-step aging treatment induces dense nanoscale precipitation of two interspersed precipitate phases, namely H-phase and a new cubic Ni-rich precipitate phase, resulting in the highest hardness exhibited yet by this family alloys. The composition of cubic Ni-rich precipitates was measured using atom probe tomography to be approximately $Ni_{61.5}Ti_{31}Hf_{7.5}$, while HAADF-STEM revealed a 54 atom motif cubic structure (a= 8.87 Å ), and electron diffraction showed that the structure belongs to the $pm\bar{3}m$ (No. 221) space group.

* Present Address: Mechanical Engineering and Materials Science and Engineering, Georgia Institute of Technology, Atlanta, GA 30332 USA
† Corresponding Author: baminahmadi@mines.edu

1.  **Introduction**

High-hardness, binary nickel titanium (NiTi) of compositions within the range of 54 – 56 at.% nickel have attractive attributes for tooling, wear, and other tribological applications that are equal to and even beyond the capabilities of steel and ceramic bearing materials [1–6]. Unique features of these alloys include corrosion resistance, cavitation erosion resistance, nonmagnetic behavior and high hardness, while being less dense and more resistant to denting damage compared to tool steels [3,7,8]. Recent reports on these alloys show excellent tribological properties under oil-lubricated rolling and sliding contact applications [4,9,10]. Moreover, combined high hardness and low effective modulus of these alloys [7,8,11] allows them to elastically accommodate large amounts of deformation by distributing loads to a larger contact area, thereby mitigating plastic strain accumulation, while maintaining comparable rolling/sliding performance to high performance steels [3–5]. Therefore, very Ni-rich NiTi alloys are drawing attention for advanced structural applications such as bearing materials for the water recycling system and environmental control system in the International Space Station [2].

High hardness in nickel-rich binary NiTi alloys is mainly attributed to nanoscale $Ni_4Ti_3$

precipitates [12–16] and to a lesser degree anti-site defects or Ni clustering due to excess Ni atoms within the matrix [17–19]. Moreover, coherency strain fields surrounding the $Ni_4Ti_3$ precipitates within the matrix can further enhance the hardness [13,20]. While coherency strains around $Ni_4Ti_3$ precipitates in 54 – 56 at.% Ni-containing NiTi alloys have not been quantitatively determined, Tirry et al. [20] measured up to 2% coherency strains around $Ni_4Ti_3$ precipitates in a $Ni_{51}Ti_{49}$ alloy. $Ni_4Ti_3$ precipitates readily form in very Ni-rich NiTi [13,14] due to similar stoichiometry with the bulk alloys (~57 at.% Ni) and a strong driving force for precipitation and growth resulting from the difference in free energy between the supersaturated NiTi matrix phase and the equilibrium B2 NiTi matrix containing the $Ni_4Ti_3$ phase [16]. Therefore, $Ni_4Ti_3$ will rapidly precipitate during cooling from solid-solution temperatures, making it nearly impossible to suppress the precipitation in bulk samples using typical quenching processes [13,14]. Solution treating Ni-rich binary NiTi alloys in the single phase B2 region is key to maximizing the hardness of these alloys [16]. But aging after solution treatment does little to further improve hardness and is often undesirable promoting the over-coarsening and decomposition of $Ni_4Ti_3$ into incoherent $Ni_3Ti_2$ and $Ni_3Ti$ phases, resulting in a loss in hardness [13]. Recent reports suggest that adding large ternary solute atoms to NiTi alloys such as Nb, Mo, and Hf slows the precipitation kinetics [21–23] and therefore, slows or even prevents $Ni_4Ti_3$ over-coarsening and/or decomposition into undesirable phases during aging.

Benefits of ternary alloying to develop improved NiTi-alloys for tribological applications were first reported by Dellacorte et al. [1] who observed improved strength and wear resistance by substituting 1 at.% Hf for Ni ($Ni_{54}Ti_{45}Hf_1$ vs. $Ni_{55}Ti_{45}$). The vacuum induction melted (VIM) and solution hardened (1000ºC (2 h)wQ) $Ni_{54}Ti_{45}Hf_1$ alloy microstructure was more uniform and free of visible flaws compared to the more meticulously prepared powder metallurgy (PM) processed Ni$_{55}$Ti$_{45}$ samples, which resulted in as good or better rolling contact field (RCF) performance and fewer sporadic failures [1]. Further *in-situ* deformation studies by Casalena et al. [24] revealed that 1% Hf addition enabled subtle control of precipitation and matrix chemistries in Ni$_{54}$Ti$_{45}$Hf$_1$ alloy. They reported a fully recoverable pseudoelastic response of 4% under uniaxial compression which was accompanied by near-zero hysteresis in vacuum induction melted (VIM) and prepared powder metallurgy (PM) specimens. This remarkable mechanical behavior lead to a 40% increase in toughness compared to Ni$_{55}$Ti$_{45}$ alloy [24]. Moreover, Khanlari et al. [25] reported that Ni$_{54}$Ti$_{45}$Hf$_1$ performed better than Ni$_{55}$Ti$_{45}$ during oil lubricated reciprocating sliding wear tests and, in particular, exhibited better resistance to subsurface crack initiation, though understanding of the connections between the improved tribological performance and the composition or microstructure was lacking. Further examination of NiTiHf tribological alloys by Hornbuckle et al. [21] showed the influence of subtle changes in Hf content (1 – 4 at. %) in Ni rich (52 – 56 at. %) NiTi alloys on hardness. It was reported that the Hf-containing alloys exhibited higher hardness compared to corresponding binary NiTi alloys with the same Ni content [21]. Bearing level hardness (679 HV) was achieved in Ni$_{56}$Ti$_{40}$Hf$_4$ (highest Hf content in their study) when the alloy was solution annealed and aged at 400 ºC for up to 300 h [21]. The increase of hardness was attributed to formation of both Ni$_4$Ti$_3$ and H-phase precipitates [21] combined with a Hf solid-solution strengthening effect due to its larger atomic size [26].

H-phase precipitation was initially observed in a high Hf (15 at. %) NiTiHf alloy [27]. More recently, in studying moderate Hf (6 – 9 at. %) NiTiHf alloys, Amin-Ahmadi et al. [28] reported that strain fields up to 2.5% form near coherent H-phase precipitates and further enhance resistance to plastic deformation, which can be inferred to also lead to increases in hardness. A related report by Amin-Ahmadi et al. [29] showed that in low Hf (< 9 at.%) $Ni_{50.3}Ti_{40.7+x}Hf_{9-x}$ alloys, traditional heat treatment (i.e. solution treatment followed by typical aging at 550 °C) led to heterogeneous nucleation of H-phase precipitates along grain boundaries resulting in poor mechanical performance. However, adding an intermediate preaging step (300 °C for 12 h) followed by final aging at 550 °C induced homogenous nucleation and consequently, a dense distribution of H-phase within the matrix that improved functional properties in the material [29]. The effect of this type of preaging step on more Ni-rich NiTiHf alloys is unknown. Moreover, the few NiTiHf tribological alloy compositions investigated only contain low levels of Hf (up to 4 at.% [21]).

The objective of this study is to develop new NiTiHf alloys with the intent to enhance their tribological performance compared to $Ni_{54}Ti_{45}Hf_1$ alloy and the alloys explored by Hornbuckle et. al [21]. Ni content was varied across the range known to be better suited for tribological sustainability (54 – 56 at.%). The Hf content was varied between the range that would not change the precipitation phases (1 – 3 at.% Hf still allows for primary $Ni_4Ti_3$ precipitation as the primary strengthening phase and slows the kinetics [30]) and the content that was expected to change the primary precipitation phase (8 at.% Hf is known to result in H-phase precipitation for alloys with other Ni contents up to 51.5 at.% [29]) to assess whether increased tribological properties can be attained. If this was indeed the case, we intended to understand why by examining changes in microstructures, with an emphasis on precipitate phases and their morphologies [30]. Hardness was used as a 1st-order screening metric for improved tribological properties and it was found that $Ni_{56}Ti_{36}Hf_8$ and $Ni_{56}Ti_{41}Hf_3$ alloys exhibited the highest hardness values and in turn, the best improvements in RCF performances via abilities to sustain higher contact stresses [30]. Here, we proceed to present our detailed findings on why these alloy-heat treatment combinations result in higher hardness through detailed transmission electron microscopy (TEM) studies of different heat treatment strategies that provide understanding to the hardness improvements, and why the two alloy chemistries respond better to unique heat treatment strategies. In a companion work, we then examine the post-rolling contact fatigue deformation behavior and elucidate the deformation mechanisms that led to RCF performance improvements [31].

## 2. Methods

### 2.1. Materials and processing

The $Ni_{56}Ti_{41}Hf_3$ and $Ni_{56}Ti_{36}Hf_8$ (at.%) alloys (target compositions) were vacuum-induction-melted (VIM) using a graphite crucible and cast into ingots 30 mm in diameter and 600 mm long. The ingots were homogenized at 1050 °C for 24 h, then sealed inside mild steel cans and hot-extruded into 11.5 mm diameter rods at 1000 °C. Chemical analysis of the extruded rods by inductively coupled plasma atomic-emission spectroscopy confirmed that the measured composition for the $Ni_{56}Ti_{41}Hf_3$ alloys was 56.56 at% Ni, 40.44 at.% Ti, 2.97 at.% Hf and 0.03 at.% Zr. The measured composition for the $Ni_{56}Ti_{36}Hf_8$ alloy was 56.47 at% Ni, 35.57 at.% Ti, 7.88 at.% Hf and 0.08 at.% Zr. These compositional measurements are within experimental error of the aim compositions. Samples sectioned from the extruded rods were vacuum encapsulated in quartz tubes under Ar and solution-annealed at 1050 ºC for 0.5 h and water quenched, followed by preaging at 300 ºC for 12 h, then final aging at 550 ºC for 4 h. Other test samples were directly aged at 550 ºC for 4 h after the solution annealing step (without preaging). Both types of heat treatments are schematically illustrated in Fig 1(a).

Vickers hardness testing using a LECO LM series digital microindentation hardness tester was conducted on all samples after being polished through 1200 grit SiC paper. Ten indents were performed at random locations and the average value and corresponding standard deviation are reported. Indents were made at least 5 indent lengths away from each other to prevent inconsistent results due to work hardening around the indents.

*2.2. Transmission Electron Microscopy*

Conventional bright-field TEM (BF-TEM), selected area electron diffraction (SAED) and high-resolution TEM (HRTEM) were performed for the majority of the microstructural characterization using a FEI Talos TEM (FEG, 200 kV). High resolution high-angle annular dark-field scanning TEM (HR HAADF STEM) was performed to achieve atomic resolution on the peak hardened $Ni_{56}Ti_{36}Hf_8$ condition (Fig. 9(c)) using a FEI Titan 80-300 with electron-probe Cs-correction operated at 300 kV. The TEM foils were prepared by thinning specimens to 100 µm using 1200 grit SiC paper and punching out 3 mm disk samples. Each sample was then electro-polished in an electrolyte solution of 30% $HNO_3$ and 70% methanol (by volume) at around -35 ºC using a current of 20 mA at 10 V.

To measure the size of various precipitates and their interparticle distance (the edge-to-edge distance between precipitates), several HRTEM images were taken from various regions. This measurement was repeated for more than 100 precipitates on each sample and average precipitate size, average interparticle distance and their corresponding standard deviation is reported. Note that the interparticle distances were measured from 2D HRTEM images. Since these measurements can be influenced by the thickness of the TEM foil, the regions with similar thickness were chosen.

To quantify the strain fields in HRTEM images, the geometrical phase analysis (GPA)

technique was used. GPA is an image-processing technique sensitive to small displacements of the lattice fringes in HRTEM images [32]. A Gaussian selection window was used in the GPA analysis with a diameter such that the spatial resolution of strain determination was 2 nm.

2.3. *X-ray Diffraction*

Lattice parameters of the B2 and $Ni_4Ti_3$ precipitate phases in the $Ni_{56}Ti_{41}Hf_3$ alloy were experimentally measured via high-energy (E = 80.72 keV) wide-angle x-ray scattering (WAXS) experiments performed at the 1-ID beamline of the Advanced Photon Source (APS), Argonne National Laboratory (ANL). Cylindrical samples (2mm(d) x 4mm(l)) were centerless ground from the extruded rod described in Section 2.1. The samples (vacuum encapsulated in a quartz tube under Ar) were solution-annealed at 1050 ºC for 0.5 h and water quenched, followed by preaging at 300 ºC for 12 h, and final aging at 550 ºC for 4 h.

The beam was vertically focused using refractive lenses to ~ 1.5 µm (full width half maximum) while the horizontal size was defined to 100 µm using slits. A 2048 × 2048 area detector (GE Angio) with 200 × 200 µm$_2$ pixels was placed 2250 mm downstream from the sample. Diffraction patterns of a NIST standard Cerium Oxide powder were used to calibrate the detector distance and tilt angles. Ten 1 second detector exposures were first corrected by subtracting a "dark" image (image taken with the beam turned off) and then summed.

Rietveld refinement [33] of the resulting data was performed using GSAS-II [34]. The background was fit using a 2$_{nd}$ order Chebyshev polynomial. The lattice parameter of the B2 structure was refined using the $Pm\bar{3}m$ space group and an initial lattice parameter value of a = 2.995 Å. Then, the $Ni_4Ti_3$ phase was introduced, and its lattice parameters and phase fraction were refined together with the B2 lattice parameter and phase fraction, using the $R\bar{3}r$ space group and initial lattice parameters values a = 6.7729 Å, $\alpha$=113.861° and an initial phase fraction of 17%. Some texture was observed; 6th order spherical harmonic expansions were then used to fit the texture of each phase, to ensure it did not bias the lattice parameter determination.

### 2.4. Atom Probe Tomography

3D atom probe tomography (APT) was performed on a needle using a CAMECA LEAP 4000 XSi lazer pulsed atom probe tomography unit to determine the composition of the cubic Ni-rich and H-phase precipitates in more detail within a $Ni_{56}Ti_{36}Hf_8$ sample after the three-step peak aging treatment (see Fig. 9(b)). The needle sample (~80 nm diameter × ~80 nm height) was prepared using focused ion beam (FIB) milling in a FEI Helios 600i dual beam microscope. Final specimen radii varied between 50 and 100 nm after a 2 kV cleaning procedure to reduce gallium ion damage on the surface. To measure the composition, a mass spectroscopy measurement was performed and from these spectra, a proxigram was prepared for each precipitate type in addition to the bulk matrix material. The mass-to-charge ratio was ranged to full width tenth maximum and smaller peaks were ranged to background.

## 3. Results

### 3.1. Micro-hardness responses to heat treatments

**Fig. 1**(a) shows the temperature and time profiles of the 2-step and 3-step heat treatments performed on both NiTiHf alloys. The corresponding average Vickers micro-hardness values after each of the steps for these two treatments are presented in **Fig. 1**(b), with errorbars representing +/- 1 standard deviation. Conventional aging of the $Ni_{56}Ti_{41}Hf_3$ alloy (550 °C for 4 h) results in a slight decrease in hardness (to 682 HV) compared to the solution annealed material (707 HV), while the preaging treatment (300 °C for 12 h), by itself, results in the maximum

hardness observed for the alloy (752 HV). Preaging of this alloy followed by conventional aging reduced the hardness back to the solution-annealed level.

Aging of the $Ni_{56}Ti_{36}Hf_8$ alloy at either 550 °C for 4 h (700 HV) or 300 °C for 12 h (705) had a relatively small but negative effect on hardness compared to the solution annealed condition (716 HV). In this case, the maximum hardness was observed after preaging followed by aging at 550 °C for 4 h (769 HV). But using this 3-step process and extending the aging time at 550 °C from 4 h to 12 h resulted in a decrease in hardness back to the solution annealed level.

### 3.2. Microstructure responses to heat treatments

Table 1**Error! Reference source not found.** summarizes the observed precipitate phases and their morphologies for the $Ni_{56}Ti_{41}Hf_3$ and $Ni_{56}Ti_{36}Hf_8$ alloys subjected to each of the different heat treatment steps, together with the corresponding hardness value measured for each bulk sample. The following subsections proceed to describe in more detail the microstructures specific to each heat treatment, and its impact on the hardness of the two Ni-rich NiTiHf alloys.

#### 3.2.1. Solution annealing

**Fig. 2**(a) is a BF-TEM micrograph of $Ni_{56}Ti_{41}Hf_3$ after solution annealing (SA) at 1050 °C for 0.5 h, followed by a water quench. The corresponding SAED pattern is also shown in the bottom left inset. The main spots in the SAED pattern belong to the B2 cubic austenite structure and the super reflections along $1/7<321>_{B2}$, indicated by dashed lines, originated from two different variants of $Ni_4Ti_3$ precipitates. **Fig. 2**(b) is a HRTEM micrograph taken along $[111]_{B2}$ // $[111]_R$ zone axis (where R refers to the rhombohedral structure of the $Ni_4Ti_3$ precipitates). The $Ni_4Ti_3$ precipitates are trigonal structure, but crystallographically they are indexed using the rhombohedral setting of the R3 space group. Note, even though these precipitates can be indexed

as rhombohedral, these are not the "R-phase" observed in near-equiatomic NiTi alloys [35,36]. The morphology of a monolithic $Ni_4Ti_3$ precipitate surrounded by B2 matrix is outlined with a dashed line, and the corresponding fast Fourier transform (FFT) pattern is shown in the bottom left inset. The precipitates have "equiaxed" morphology, which is different from the reported lenticular morphology for $Ni_4Ti_3$ precipitates in binary NiTi [14,15,27]. The average size of these $Ni_4Ti_3$ precipitates is 51 ± 18 nm (length and width) and their average interparticle distance is 26 ± 9 nm.

The microstructure of $Ni_{56}Ti_{36}Hf_8$ after solution annealing at 1050 °C for 0.5 h, followed by a water quench is presented in **Fig. 3**. Precipitates were not obviously observed within the matrix in examining the BF-TEM micrograph (**Fig. 3**(a)); however, the SAED pattern taken from the lower grain shows faint super reflections along 1/3<110> (indicated by white arrowheads). These super reflections originate from H-phase precipitation [37,38]. The low intensity of super reflections in the SAED pattern suggest a low volume fraction and/or very small precipitate size. **Fig. 3**(b) shows a HRTEM image taken along $[111]_{B2}$. The FFT pattern (inset in **Fig. 3**(b)) of this image shows faint super reflections that originate from the H-phase precipitate (indicated by black arrowheads), confirming the presence of a single H-phase precipitate. Together these results indicate that fine H-phase precipitates (2-5 nm in diameter) formed randomly throughout the matrix in the $Ni_{56}Ti_{36}Hf_8$ alloy after water quenching from the solution annealing temperature.

*3.2.2. Solution annealing and aging at 550 ºC for 4 h*

The microstructure of $Ni_{56}Ti_{41}Hf_3$ after solution annealing and subsequent aging at 550 ºC for 4 h is presented in **Fig. 4**. The BF micrograph (**Fig. 4**(a)) and its corresponding SAED pattern (inset) show that the microstructure has a "blocky" or mosaic morphology consisting of $Ni_4Ti_3$ precipitates within a B2 matrix. The average $Ni_4Ti_3$ precipitate length and interparticle distance is $109 \pm 33$ nm and $92 \pm 24$ nm, respectively.

**Fig. 4**(b) is a HRTEM micrograph of a blocky $Ni_4Ti_3$ precipitate surrounded by the B2 matrix. It is interesting to note that the precipitate is actually composed of two coalesced $Ni_4Ti_3$ variants. The corresponding FFT pattern ($[111]_R$/ /$[111]_{B2}$) shown in the inset of **Fig. 4**(b) was taken from the boundary region of the two $Ni_4Ti_3$ variants (indicated by the white dashed box in **Fig. 4**(b)) and clearly indicates a twin relationship. The twin plane is indexed as $(\bar{1}5\bar{4})_R$. **Fig. 4**(c) shows the local **g**-map taken using $\mathbf{g} = (\bar{1}01)_R$ and $\mathbf{g} = (\bar{1}01)_R$ from the two variants of $Ni_4Ti_3$, (indicated on the FFT pattern in **Fig. 4**(b) by a white circle). The arrangement of misfit dislocations (shown as hot spots in **Fig. 4**(c)) confirms the curved shape of the boundary. The curved nature of the twin interface in $Ni_4Ti_3$ has been also reported by Tadaki et al. [39] in binary NiTi alloys of lower Ni content and is attributed to the formation of a twin relation using equivalent $\{154\}_R$ type planes [39]. Therefore, the boundary typically lies on multiple unique planes that falls in the same family, effectively changing in direction and causing a curvature along the interface. Inverse FFT filtering was also performed on a part of the interface to identify the atomic arrangement along the interface and is shown in **Fig. 4**(d). The magnified interface is outlined by its repetitive structural units (white polygons) showing the arrangement and symmetry of atoms along the boundary that is shared between the two crystals. The coincident site lattice (CSL) boundary was detected along the interface with a 22° misorientation angle along the $[111]_R$ axis, which is consistent with the misorientation angle for $\Sigma = 7$ type boundary along $[001]_{HCP}$ // $[111]_R$ axis [40,41].

**Fig. 5** depicts the microstructure of Ni$_{56}$Ti$_{36}$Hf$_8$ after 550 ºC for 4 h aging of the solution annealed material. The BF-TEM image and corresponding SAED pattern (inset) in **Fig. 5**(a) show the presence of nanoscale H-phase precipitates within the bulk of the B2 matrix. While the majority of the microstructure consists of a dense homogeneous distribution (49% area fraction) of nanoscale H-phase precipitates in a B2 matrix. HRTEM investigation of the matrix (Fig. 5(d)) shows that the H-phase precipitates are ellipsoidal in shape with average dimensions of 21 ± 6 nm (length) and 8 ± 2 nm (width); the interparticle distance was 17 ± 6 nm. The H-phase precipitates in this study has a similar morphology to that generally reported for H-phase in NiTiHf alloys [27,28,37,38]. The microstructure of the same alloy after solution annealing showed only a very small fraction of very fine H-phase precipitates (**Fig. 3**(b)). Aging at 550 ºC for 4 h resulted in additional nucleation and growth of the precipitates, thus much more intense super reflections along 1/3<011> were observed in the SAED pattern. Ni$_4$Ti$_3$ precipitates were not observed in this sample.

**Fig. 5**(a) also reveals the presence of a heterogeneously nucleated phase in the vicinity of HfO$_2$ particles. Electron Dispersive Spectroscopy (EDS) analysis on these precipitates indicated that they were comprised of 65 ± 4 at.% Ni and 35 ± 2 at.% Ti, which is nominally Ni$_3$Ti$_2$, however, no Hf was detected. The size of these precipitates is 648 ± 277 nm (length), 191 ± 37 nm (width) and they are spaced 673 ± 271 nm apart. Corresponding SAED patterns (Fig. 5 (b,c)) from [201]$_{Ni_3Ti_2}$ and [001]$_{Ni_3Ti_2}$ zone axes confirm the orthorhombic structure of Ni$_3$Ti$_2$ phase. HRTEM micrograph (Fig. 5(d)) taken along [111]$_{B2}$ shows a large Ni$_3$Ti$_2$ phase and H-phase precipitates within B2 matrix. Corresponding FFT pattern of Ni$_3$Ti$_2$ phase in Fig. 5(d) and an additional HRTEM image (not shown here) is indexed as [501]$_{Ni_3Ti_2}$ and [201]$_{Ni_3Ti_2}$ zone axes of orthorombic structure. These precipitates appear to have mostly nucleated heterogeneously

around the larger HfO$_2$ particles most likely due to local Hf depletion, and perhaps further encouraged by stress concentrations about the HfO$_2$ particles. It is known that such large heterogeneous precipitates lose their coherency and subsequently their effectiveness at halting deformation processes [42], and can also become the sources for fatigue failure, wear failure, and localized corrosion [43].

*3.2.3. Solution annealing and preaging at 300 ºC for 12 h*

To understand the role of the preaging treatment, the microstructures of the Ni$_{56}$Ti$_{36}$Hf$_8$ and Ni$_{56}$Ti$_{41}$Hf$_3$ alloys after preaging at 300 ºC for 12 h were studied. **Fig. 6**(a) is a BF-TEM micrograph and corresponding SAED pattern (inset) of Ni$_{56}$Ti$_{41}$Hf$_3$ after preaging and confirms the widespread precipitation of equiaxed Ni$_4$Ti$_3$ phase. The size of the blocky Ni$_4$Ti$_3$ precipitates is 81 ± 15 nm (length and width) and their interparticle distance is 34 ± 19. The significant diffraction contrast surrounding the precipitates in the BF-image is due to the coherency strain fields around the precipitates.

In contrast, the BF-TEM micrograph of the Ni$_{56}$Ti$_{36}$Hf$_8$ alloy after preaging (**Fig. 6**(b)) indicates a dramatically different outcome after the same preaging, with no clear precipitation within the grain. However, the corresponding SAED pattern (bottom left inset) shows faint spots corresponding to occasional (< 5 nm) H-phase precipitates similar to what was observed in the solution annealed and quenched condition (**Fig. 3**). An additional SAED pattern provided in the upper right inset was taken slightly off the [111]$_{B2}$ zone axis showing diffuse intensity patterns. The diffuse intensities appear more clearly in the higher index tilted zone axis in the form of line shapes with a periodic symmetry around the bright B2 reflections. In this case, the faint super reflections from H-phase precipitates are also observed in the pattern (top right inset). The diffuse intensities with periodic character in reciprocal space indicate the existence of short-range order in the real-space lattice as a precursor to widespread formation of precipitates. Such diffuse intensities have been reported in binary $Ni_{50.6}Ti_{49.4}$ by Pourbabak et al. [44] after low-temperature aging and were attributed to the existence of micro-domains in the form of clusters of Ni atoms as precursors to formation of $Ni_4Ti_3$ nanoprecipitates [44]. Recently, Amin-Ahmadi et al. [29] reported diffuse diffraction intensities for $Ni_{50.3}Ti_{41.2}Hf_{8.5}$ due to Hf and/or Ni atom clusters formed after low-temperature aging as a precursor to H-phase precipitation, which is the most likely explanation for the diffuse intensity observed in the inset to Fig. 6b.

*3.2.4. Solution annealing, preaging at 300 ºC for 12 h, and aging at 550 ºC for 4h*

**Fig. 7** is representative of the microstructure of $Ni_{56}Ti_{41}Hf_3$ after a three-step heat treatment consisting of solution annealing, preaging at 300 ºC for 12 h, and aging at 550 ºC for 4 h. The BF-TEM micrograph (Fig. 7(a)) and corresponding SAED pattern (inset) shows a blocky microstructure consisting of larger $Ni_4Ti_3$, and small densely packed H-phase and new cubic structured Ni-rich precipitates. Ongoing work is aimed at determining the exact atomic configuration of the new phase. Later in this section, for the $Ni_{56}Ti_{36}Hf_8$ alloy subjected to this same heat treatment which is observed to exhibit a higher density of precipitates of this new phase, we will show that this phase contains Hf and exhibits cubic symmetry and a 54-atom repeating motif. Therefore, it is not the same phase as the larger orthorhombic $Ni_3Ti_2$ phase that formed near the $HfO_2$ particles [31] (see Section 3.2.2).

There is significantly less strain contrast observed about $Ni_4Ti_3$ precipitates in this BF-TEM image (Fig. 7(a)) compared with the preaged condition of the same alloy (**Fig. 6**(a)). This observation is attributed to relaxation of the coherency strains as the precipitates coarsen during aging at 550 ºC [16,37]. The presence of other secondary precipitate phases (H-phase and new Ni-rich cubic phase) could be a confounding factor in this the loss in $Ni_4Ti_3$ / B2 coherency. In particular, this alloy consists of blocky $Ni_4Ti_3$ precipitates of approximately $138 \pm 41$ nm $\times$ $94 \pm 23$ nm in size.

HRTEM analysis (**Fig. 7**(b)) showed that the H-phase precipitates that are also present are ellipsoidal in shape, and formed within the B2 channels positioned between $Ni_4Ti_3$ precipitates, similar to previous observations by Hornbuckle et al. [21] for a $Ni_{56}Ti_{40}Hf_4$ alloy. The average size of the H-phase was $28 \pm 4$ nm (length) and $13 \pm 2$ nm (width) and the interparticle distance along the channels was $104 \pm 13$ nm. The HRTEM micrograph in **Fig. 7**(c) shows a small cubic Ni-rich precipitate bordered by B2 matrix and a much larger $Ni_4Ti_3$ precipitate. This rare occurrence of cubic Ni-rich precipitation in the $Ni_{56}Ti_{41}Hf_3$ alloy is confirmed via the corresponding FFT pattern (inset), which is consistent with the cubic structure we proceed to document in more detail for the $Ni_{56}Ti_{36}Hf_8$ alloy subjected to the same heat treatment.

The microstructure of the $Ni_{56}Ti_{36}Hf_8$ alloy after the three-step heat treatment is shown in **Fig. 8**. It is obvious from the BF-TEM micrograph (**Fig. 8**(a)) that the mottled appearance of the precipitate morphology is much finer than the blocky structure shown in **Fig. 7**(a) for the $Ni_{56}Ti_{41}Hf_3$ alloy. The microstructure consists primarily of a very dense distribution of interspersed nanosized H-phase and cubic Ni-rich precipitates. Large heterogeneous orthorhombic $Ni_3Ti_2$ phases not shown here were also observed in the vicinity of $HfO_2$ oxide particles, as in Fig. 5.

**Fig. 8**(b, c, d) are SAED patterns taken along $[001]_{B2}$, $[111]_{B2}$ and $[011]_{B2}$ zones from the region indicated by the white circle in **Fig. 8**(a)), confirming the presence of H-phase (indicated by white arrowheads) and cubic Ni-rich precipitates (indicated by red arrows in the SAEDs and by red circles in the schematics (Fig. 8(e–f)), within the B2 matrix. Since the precipitates of the cubic Ni-rich phase are too small to allow convergent beam electron diffraction methods to be used to determine their point and space groups, or EDS to measure their compositions, the following analyses were used to determine as much about this phase as we could at this time. First, the SAED patterns from the 3 zone axes were considered. It is noted that the extra reflections in the 3 major zones from the B2 phase are consistently in the 1/3 hkl positions relative to the primitive B2 phase (all reflections present). Furthermore, there are no extra reflections or extinctions in any of these zones. Since the matrix is B2, which belongs to the $Pm\bar{3}m$ space group, then the space group of the precipitates must be primitive and cubic as well given that, if it were non-cubic or some other non-primitive cubic structure, there would be extinctions in one or more of these patterns. Thus, it is concluded that the space group of the precipitates is most likely $Pm\bar{3}m$ with a lattice parameter of approximately 3 times that of the B2 parent. It is noted that the primitive cell could be one of the primitive subgroups of the $m3m$ parent phase, i.e., $P23, Pm3, P\bar{4}3m$ or $P432$ and further work is required to differentiate between these possibilities and $Pm\bar{3}m$.

Second, a HR-TEM micrograph taken along the $[001]_{B2}$ zone and corresponding FFT pattern presented in Fig. 9(a), clearly shows the dense distribution of H-phase and presence of cubic Ni-rich nanoprecipitates within B2 channels. The FFT pattern taken directly from single precipitates (Fig. 9(a)) match the SAED schematic (Fig. 8(e)) for the $[001]_P$ zone and show the expected cube-cube relationship with the B2 matrix. This phase is consistent with the "cubic $Ni_3Ti_2$" diffraction patterns reported in a binary $Ni_{52}Ti_{48}$ alloy by Karlik et al. [45]. The diffracted spots in the SAED patterns (Fig. 8(b–d)) are indexed accordingly and presented on the FFT pattern of Fig. 9(a). Using the SAED patterns, the measured lattice constant for the new phase is ~8.87 Å. Analysis of this and similar HR-TEM micrographs determined that the average size of the ellipsoidal H-phase precipitates is 23 ± 5 nm (length) and 12 ± 3 nm (width) and the interparticle distance is 7 ± 2 nm. The average size of cubic Ni-rich precipitates is 18 ± 4 nm (length) and 16 ± 5 nm (width), which mainly formed between H-phase precipitates. The size of the remaining B2 channels was very fine scale – approximately 8 ± 3 nm.

Third, analysis of APT data of a needle containing cubic Ni-rich and H-phase precipitates, part of which is visualized in Fig. 9(b), was used to assess the chemistry of the new phase. The measured overall composition of the reconstructed volume was 56.9 ± 0.4 at.% Ni, 33.9 ± 0.1 at.% Ti and 9.2 ± 0.2 at.% Hf which is slightly different from nominal composition of the alloy ($Ni_{56}Ti_{36}Hf_8$) since the APT tip does not contain any oxides / inclusions or large orthorhombic $Ni_3Ti_2$ precipitates that are also present throughout the microstructure. The plotted iso-concentration surface using the threshold value of 59 at.% for Ni and 9 at.% for Hf (Fig. 9(b)) clearly shows the existence of two types of precipitates in agreement with TEM observation (Fig. 8). The average composition of one type of precipitates is 61.5 ± 0.1 at.% Ni, 31 ± 0.3 at.% Ti and 7.5 ± 0.3 at.% Hf and the other type is 55.4 ± 0.4 at.% Ni, 32.1 ± 0.2 at.% Ti and 12.5 ± 0.4 at.% Hf. The latter composition with higher Hf content is associated with H-phase due to higher Hf content of these precipitates [37]. Therefore, the other composition is associated with the new, cubic Ni-rich phase. The diffraction spots originated from cubic Ni-rich precipitate (Fig. 8) are similar to the "cubic $Ni_3Ti_2$" reported in [45,46]; however, due to different composition (existence of Hf atoms), the phase is identified to be a new, cubic Ni-rich NiTiHf precipitate phase, which may be a NiTiHf analog to the previously reported "cubic $Ni_3Ti_2$" binary phase. The measured composition of the H-phase precipitates in this sample is very different than previously reported H-phase precipitate compositions [37]. This finding shows that the H-phase composition can vary significantly and is dependent on Ni and Hf content of the matrix. Ongoing work is necessary to elucidate the mechanisms of this variation of H-phase.

Fourth, atomic resolution HAADF-STEM analysis of the precipitate phase along [001] axis revealed a square grid-like supercell Z-contrast motif with the edges brighter than the internal atoms (Fig. 9(c)). Fast Fourier transform (FFT) (inset of Fig. 9(c)) is consistent with the SAED pattern in Fig. 8(b) and FFT in Fig. 9(a). In STEM imaging, intensity is proportional to atomic number (I $\propto Z^{1.7}$ [47]), thus columns of atoms with a higher atomic number, more Hf and/or Ni in this case, are observed along the perimeter of the large primitive cell while mixtures of Ni, Hf and Ti atoms must occupy the interior lattice sites in order to satisfy the $Ni_{61.5}Ti_{31}Hf_{7.5}$ composition. The measured supercell size in the HAADF-STEM image (Fig. 9(c)) is consistent with the lattice constant measured from SAED pattern (~8.87 Å) which further confirms the primitive nature of the cubic structure. Furthermore, considering the number of atomic columns contained in each repeating unit of the square motif, together with the space group symmetry assessed from the SAED patterns results in an understanding that the unit cell most probably consists of 54 atom sites arranged in a BCC-like pattern, ignoring elemental differences.

The use of ab initio and/or atomistic simulation techniques to help determine the exact point group and the elemental atomic site occupancies of the new cubic Ni-rich precipitate phase is challenged by the combination of the number of elements together with the size of the chemical motif – the determination of the configuration of a 54-atom unit cell composed of 3 elements present in different ratios challenges the state of the art to advance. Moreover, these results have motivated ongoing work to study the effect of bulk alloy compositional changes on the H-phase precipitate phase composition. APT results of this work showed enormous change in H-phase composition ($Ni_{55}Ti_{32}Hf_{12}$) for $Ni_{56}Ti_{36}Hf_8$ alloy compared with the APT results of H-phase composition ($Ni_{52}Ti_{19}Hf_{29}$) for $Ni_{50.3}Ti_{29.7}Hf_{20}$ alloy reported in [37], suggesting that H-phase chemistry likely varies with Ni:Hf availability from the bulk matrix.

*3.2.5. Solution annealing, preaging at 300 ºC for 12 h and aging at 550 ºC for 13h*

A three-step heat treatment with an extended age at 550ºC of 13 h was applied to the $Ni_{56}Ti_{36}Hf_8$ alloy and the resulting microstructure is shown in Fig. 10(a). The mottled 3-phase microstructure containing cubic Ni-rich precipitates and H-phase precipitates with narrow channels of B2 matrix is very similar to the microstructure observed in Fig. 8, except that the H-phase precipitates appear to have coarsened and doubled in size (Table 1) while the cubic Ni-rich precipitates are roughly the same size (Fig. 10(b)). These observations would explain the very prominent H-phase reflections and nearly invisible superlattice reflections for the cubic Ni-rich precipitates in the SAED pattern (Fig. 10(a)). The average size of the ellipsoidal H-phase precipitates is $40 \pm 8$ nm (length) and $21 \pm 6$ nm (width) and the interparticle distance is $14 \pm 4$ nm. The average size of the cubic Ni-rich precipitates is $20 \pm 4$ nm (length) and $17 \pm 3$ nm (width), which continue to reside between H-phase precipitates. The average size of the B2 channels is $15 \pm 5$ nm.

*3.2.6. $Ni_4Ti_3$ precipitate / B2 matrix coherency in $Ni_{56}Ti_{41}Hf_3$ alloy*

**Fig. 11** shows HR-TEM micrographs (a)–(c) and corresponding GPA maps (d)–(f) ($\varepsilon_{yy}$ strain component) of $Ni_{56}Ti_{41}Hf_3$ alloy samples. After solution annealing with a water quench ($SA_{WQ}$), fine $Ni_4Ti_3$ precipitates ($51 \pm 18$ nm) appear in Figs. 10 a and d to be fully coherent with the B2 matrix. To accommodate the lattice mismatch, the surrounding matrix is strained up to 1% (regions indicated by white arrows in **Fig. 11**(d)), which extends $3.1 \pm 2$ nm away from the precipitate–matrix interface. Some of the larger (50 – 80 nm) $Ni_4Ti_3$ precipitates that were inspected contain visible misfit dislocations. When the material is preaged at 300ºC (12 h) (**Fig. 11**(b)), the average size of the precipitates increases (81 ± 15 nm) and the strain fields extend as far as 10 ± 3 nm away from the precipitate–matrix interface (indicated by white arrows in **Fig. 11**(e)). The upper 2/3 of the interface highlighted in **Fig. 11**(e) is coherent; however, the lower part of the interface is beginning to lose coherency by forming misfit dislocations. The distance between these misfit dislocations was approximately 6.2 ± 2 nm (misfit dislocations are shown by white arrowheads in **Fig. 11**(e)). Further aging at 550ºC for 4 h (3-step heat treatment) (**Fig. 11**(c)), results in additional coarsening of the precipitates (138 ± 41 nm) and further loss in coherency (**Fig. 11**(f)), with misfit dislocations (shown by white triangles in **Fig. 11**(f)) spaced about 3.3 ± 1 nm apart along the interface. When the material is direct aged (SA$_{WQ}$ + 550ºC (4 h)), the precipitates exhibit a loss in coherency (similar to the 3-step heat treatment), with misfit dislocations spaced approximately 5.2 ± 2 nm apart along the precipitate–matrix interface.

## 4. Discussion

### 4.1. Strengthening mechanisms in $Ni_{56}Ti_{41}Hf_3$ and $Ni_{56}Ti_{36}Hf_8$

During the SA of the $Ni_{56}Ti_{41}Hf_3$ alloy, Ni is supersaturated in the matrix, but because of the retrograde solvus, the excess solubility for Ni in the matrix decreases with decreasing temperature and the driving force for precipitation increases to the point where, as in binary alloys, it is not possible to prevent precipitation, even with a rapid quench [13,14]. This excess Ni is removed from the matrix through formation of the metastable $Ni_4Ti_3$ phase. Similarly, in the $Ni_{56}Ti_{41}Hf_3$ alloy, $Ni_4Ti_3$ is formed even during the water quench. The high hardness (707 HV) of the alloy in the SA condition is heavily dependent on the nanoscale size (51 nm) of mostly coherent and equiaxed $Ni_4Ti_3$ precipitates with 54% area fraction.

When the $Ni_{56}Ti_{41}Hf_3$ alloy is aged at 550 °C for 4 h, the $Ni_4Ti_3$ precipitate size increases 214% in length and 171% in width compared with the SA condition and the area fraction of $Ni_4Ti_3$ increases to 61%. The combination of over-coarsened blocky $Ni_4Ti_3$ precipitates and subsequent loss in coherency strains within the B2 matrix (as evidenced by reduced strain contrast surrounding the precipitates in **Fig. 4**(b) and formation of misfit dislocations along the relaxed interface) leads to a decrease in the measured hardness to 682 HV compared with the SA condition (707 HV). However, preaging the $Ni_{56}Ti_{41}Hf_3$ alloy at 300 °C for 12 h after the SA condition (**Fig. 6**) resulted in an increase in hardness to 752 HV. This increase in hardness can be explained by the controlled increase in size of the equiaxed, semi-coherent $Ni_4Ti_3$ precipitates (81 nm), which achieve a maximum fraction in this condition (71% area fraction). Furthermore, the coherency strains have expanded further into the surrounding matrix (up to 10 nm), contributing to the hardness, even though part of the interface loses its coherency (**Fig. 11**(e)). Qualitatively, high strain contrast in the form of darker matrix regions in the BF micrograph of **Fig. 6**(a) implies that the interface is still mostly coherent. Applying a final aging step at 550 °C for 4 h, after preaging, resulted in a decrease in hardness to 710 HV mainly due to over coarsening of the $Ni_4Ti_3$ precipitates (138 nm length × 94 nm width), and a much greater loss in coherency, and a decrease in their area fraction to 63%.

The $Ni_{56}Ti_{36}Hf_8$ alloy in the SA condition (**Fig. 3**) is composed almost entirely of B2 matrix with the exception of a few H-phase precipitates, suggesting that nearly all Hf and Ni is still in solid solution. Interestingly, the hardness of SA $Ni_{56}Ti_{36}Hf_8$ (716 HV) is comparable to the hardness of SA $Ni_{56}Ti_{41}Hf_3$ alloy (707 HV) containing 51% $Ni_4Ti_3$ precipitate phase, which suggests that Hf solid-solution hardening and/or clustering of Hf and Ni atoms, plays an important role in enhancing the hardness of this particular alloy.

The hardness decreased to 700 HV after aging the Ni$_{56}$Ti$_{36}$Hf$_8$ alloy at 550 ºC for 4 h compared with 716 HV for SA condition. The majority of the microstructure in the aged alloy (**Fig. 5**) contains fine homogenous H-phase precipitates (49% area fraction), in addition to rare large island precipitates. H-phase precipitates are rich in Ni and Hf [27,37,48]; therefore, their formation depletes excess Hf and Ni from the matrix, thus the effect of solid-solution strengthening decreases. In addition to formation of the nanosized H-phase precipitates (49% area fraction) with 21 ± 6 nm (length), 8 ± 2 nm (width), the hardness slightly declined by 2.2% from the SA condition, which implies that H-phase precipitation hardening failed to compensate for the decrease of hardness due to the loss in solid-solution strengthening.

Applying a preaging step at 300 ºC for 12 h to the Ni$_{56}$Ti$_{36}$Hf$_8$ alloy (**Fig. 6**(b)) resulted in no significant changes to the microstructure compared to the SA condition (**Fig. 3**) implying that heat treating at 300 ºC for 12 h did not generate the required energy to activate widespread nucleation and/or additional precipitate growth. Consequently, the hardness (705 HV) did not change much compared to the solution annealed condition (716 HV), suggesting that solid solution hardening remains dominant.

The benefit of the preaging step on Ni$_{56}$Ti$_{36}$Hf$_8$ alloy hardness is realized only after the additional aging step at 550 ºC for 4 h (Fig. 8) when both H-phase and cubic Ni-rich precipitates nucleate from Ni-rich micro-domains resulting in superior hardness (769 HV). In the 3-step heat treated Ni$_{56}$Ti$_{36}$Hf$_8$ alloy, widespread nucleation and hindered growth of the dense interspersed precipitates (54% H-phase area fraction and 33% cubic Ni-rich precipitate area fraction) restricts the remaining B2 matrix to ultra-narrow (~ 8 nm) channels and thus the lowest observed fraction of available matrix (23% area fraction) of any condition in this study. Therefore, the hardness reached 769 HV, a 7.4% and 9% increase compared to the SA and preaged conditions, respectively. In contrast, in the $Ni_{56}Ti_{41}Hf_3$ alloy, there was only rare cubic Ni-rich precipitates that nucleated adjacent to the $Ni_4Ti_3$.

Comparing maximum hardness for the two alloys, the highest hardness observed in the $Ni_{56}Ti_{41}Hf_3$ alloy corresponded to high volume fraction (71%) of relatively coarse, but semi-coherent $Ni_4Ti_3$ phase, which occurred in the SA plus aged condition. The maximum hardness in the $Ni_{56}Ti_{36}Hf_8$ alloy occurred with a mixed microstructure of H-phase and cubic Ni-rich precipitates that interacted to keep both phases relatively fine and together resulted in a volume fraction of ~87% coherent strengthening phase. The strengthening effect of H-phase is well documented [28,49,50]. The coherency strains associated with the cube-to-cube orientation relationship between the B2 matrix and the cubic Ni-rich precipitates likely harden the material as well, and between the two coherent precipitate phases would make dislocation motion extremely difficult. Together, the two phases interact, preventing either from coarsening too much during the 550 ºC / 4 h age, resulting in an extremely high total volume fraction of fine coherent strengthening phase.

After further aging of the $Ni_{56}Ti_{36}Hf_8$ alloy at 550 ºC for 13 h, cubic Ni-rich precipitates, H-phase and the B2 matrix are preserved (Fig. 10). However, the H-phase precipitates coarsen, resulting in an increased fraction (61 % area fraction) and the B2 channels broaden at the expense of the cubic Ni-rich precipitates (19 % from 33 % area fraction), which leads to a decrease in the hardness (718 HV).

*4.2. Ni₄Ti₃ coarsening behavior*

As discussed in Section 3.5.1. the size, distribution, and coherency strain fields of Ni₄Ti₃ precipitates play an important role in the overall hardness of NiTi-based alloys. It is known that the Ni₄Ti₃ precipitates in binary NiTi tend to be lenticular in shape due to lattice mismatch differences with the matrix [22]. The lattice mismatch surrounding coherent Ni₄Ti₃ precipitates is quantified by measuring the orientation correspondence between rhombohedral (Ni₄Ti₃) and cubic (matrix) unit cells. The habit plane of the Ni₄Ti₃ precipitate is {111}$_{B2}$ and the orientation relationship is (001)$_R$//(111)$_{B2}$, [010]$_R$//[$\bar{2}\bar{1}3$]$_{B2}$ [39,51]. Therefore, the lattice parameter of the B2 ($a_{B2}$) and the unit diagonal dimension of the of the rhombohedral unit cell ($\sqrt{3}c_r$) should be in close agreement with one another (e.g., $c_R$ = [111]$_{B2}$). Given the lattice parameter of the rhombohedral structured Ni₄Ti₃ unit cell ($c_R$ = 6.70 Å [39]) and B2 austenite unit cell ($a_{B2}$ = 3.01 Å [39]) in binary NiTi, the precipitate / matrix correspondence is calculated for a specified direction via dot product and the matrix tensor for the Ni₄Ti₃ unit cell. Equation 1 quantifies the lattice mismatch between the unit diagonal rhombohedral unit cell and the edge length of the B2 unit cell along direction normal to the habit plane:

$$\text{Ni}_4\text{Ti}_3 \text{ // B2 lattice mismatch} = \frac{\sqrt{3}c_R - \sqrt{3}a_{B2}}{\sqrt{3} * a_{B2}} * 100 \qquad (1)$$

Lattice mismatch calculation along the direction normal to the habit plane shows 2.9% contraction in the precipitate with respect to the matrix [20,52], which causes the most suppressed direction of growth for Ni₄Ti₃. Conversely, the directions in the habit plane (normal to the direction with largest mismatch) contain a precipitate/matrix mismatch of 0.5% [39] and, therefore, much less strain [53]. This explains why in the directions of the habit plane, the precipitate dimensions are much larger than in the habit plane normal direction, resulting in a lens-shaped precipitate morphology [22,53]. Therefore, the aspect ratio of the $Ni_4Ti_3$ precipitates reported in Ni-rich binary NiTi alloy compositions is 5 – 10.5 [39,42,54].

However, the aspect ratio of $Ni_4Ti_3$ precipitates in this study is dramatically different, with a value of about 1 – 1.5. Moreover, the similarities in shape and aspect ratio of two different variants that are visible from the $[111]_{B2}$ zone suggests that the precipitates grow in an equiaxed shape in 3 dimensions. To explain the morphology change of the precipitate to equiaxed morphology in the $Ni_{56}Ti_{41}Hf_3$ alloy as a result of the lattice mismatch in the direction normal to the habit plane (Equation 1), lattice parameters were determined from high-energy wide-angle x-ray scattering (WAXS) to be a = 6.722(6) Å, $\alpha = 113.866(4)°$ for $Ni_4Ti_3$ and a = 2.967(6) Å for B2. Therefore, the lattice mismatch in the direction normal to the habit plane is 1.1% (Equation 1). This notable decrease (from 2.9%) leads to growth of precipitates along directions normal to habit plane in addition to the habit plane direction and thus explains the equiaxed morphology of the $Ni_4Ti_3$ (**Fig. 2**) that is rather unique to very Ni-rich (52 – 56 at. % Ni) ternary NiTi-X alloys [22,55].

## 5. Conclusions

We reported upon understanding the heat-treatment – hardness – microstructure relationships of two new NiTiHf alloys known to exhibit high hardness and superior tribological properties relative to previously developed NiTi and NiTiHf alloys [30]. Overarching structure-property mechanisms in this material were discussed to explain how existence of various precipitates phases such as $Ni_4Ti_3$, a cubic Ni-rich precipitate, and H-phase within the B2 matrix enhance hardness. The main findings can be summarized as follows.

- Blocky $Ni_4Ti_3$ precipitation is the dominant hardening mechanism in the $Ni_{56}Ti_{41}Hf_3$ alloy.

The highest hardness for $Ni_{56}Ti_{41}Hf_3$ alloy was observed after preaging ($SA_{WQ}$ + 300 °C/ 12 h) where the fraction of $Ni_4Ti_3$ was maximized and the coherency strains had the largest impact on the matrix. Aging at higher temperature (550°C/4 h) lead to a reduction in the hardness due to over-coarsening of the $Ni_4Ti_3$ and loss of coherency.

- In $Ni_{56}Ti_{41}Hf_3$, the blocky shape of the $Ni_4Ti_3$ is due to reduced lattice mismatch between the B2 matrix and the $Ni_4Ti_3$ precipitate along the direction normal to the habit plane, allowing the precipitate phase to grow equally in directions normal and parallel to the habit plane.

- A Ni/Hf solid-solution effect or clustering of these atoms plays an important role in the hardness of the SA $Ni_{56}Ti_{36}Hf_8$ alloy, which achieved a comparable hardness to the $Ni_4Ti_3$ precipitate strengthened SA $Ni_{56}Ti_{41}Hf_3$ alloy.

- The highest hardness was achieved in the $Ni_{56}Ti_{36}Hf_8$ alloy after applying an intermediate ageing step (300°C for 12 h) followed by normal aging (550 °C for 4 h). This alloy contained a high volume fraction of intermixed cubic Ni-rich and H-phase precipitates. Together these two phases resulted in nearly 87% fine, coherent, precipitate phases in a B2 matrix that consisted of only fine nanometer sized channels.

- In the $Ni_{56}Ti_{36}Hf_8$ alloy under 3-step heat treatment, the the cubic Ni-rich precipitates were determined to be a new cubic NiTiHf phase with a composition of approximately $Ni_{61.5}Ti_{31}Hf_{7.5}$, with a structure belonging to the $p m \bar{3} m$ (No. 221) space group with a= 8.87 Å, a 54-atom motif defined by chemical segregation of heavier elements (Hf and/or Ni) to the edges of the unit cells, and having a cube– to – cube lattice correspondence with the B2 austenite matrix.

- Through APT of the $Ni_{56}Ti_{36}Hf_8$ alloy under 3-step heat treatment, it was revealed that the

composition of the H-phase is very different than the previously measured composition within a Ni$_{50.3}$Ti$_x$Hf$_{20}$ alloy. This shows that the H-phase composition can vary, depending on the Ni and Hf compositions of the matrix. Exact mechanisms and/or chemical models of this variation of H-phase is the focus of ongoing work.

**Acknowledgements**

This work was conducted within the National Science Foundation (NSF) I/UCRC Center for Advanced Non-Ferrous Structural Alloys (CANFSA), which is a joint industry-university center between the Colorado School of Mines and Iowa State University. Additional support was provided for this work through the NASA Transformative Aeronautics Concepts Program (TACP), Transformational Tools & Technologies Project under the guidance of Othmane Benafan, Technical Lead for Shape Memory Alloys. The HAADF-STEM imaging was performed at the Center for Electron Microscopy and Analysis (CEMAS) at the Ohio State University. The APT measurements were conducted by David Diercks at the Colorado School of Mines. This research used resources of the Advanced Photon Source, a U.S. Department of Energy (DOE) Office of Science User Facility operated for the DOE Office of Science by Argonne National Laboratory under Contract No. DE-AC02-06CH11357. The X-ray diffraction experiments were conducted at ANL, APS, 1-ID beamline.

**References**

[1] C. Della Corte, M.K. Stanford, T.R. Jett, Rolling contact fatigue of superelastic intermetallic materials (SIM) for Use as resilient corrosion resistant bearings, Tribol. Lett. 57 (2015) 1–10.

[2] C. DellaCorte, W.A. Wozniak, Design and Manufacturing Considerations for Shockproof and Corrosion-Immune Superelastic Nickel-Titanium Bearings for a Space Station Application, 2012.

[3] C. DellaCorte, R.D. Noebe, M.K. Stanford, S.A. Padula, Resilient and Corrosion-Proof Rolling Element Bearings Made from Superelastic Ni-Ti Alloys for Aerospace Mechanism Applications, Rolling Element Bearings, 9th Volume, STP 1542, Yoshimi R. Takeuchi and William F. Mandler, Editors, ASTM International, West Conshohocken, PA (2012), pp. 143-166.

[4] C. Dellacorte, S. V Pepper, R. Noebe, D.R. Hull, G. Glennon, Intermetallic nickel-titanium alloys for oil-lubricated bearing applications, NASA/TM—2009-215646 (March 2009).

[5] P. Clayton, Tribological behavior of a titanium-nickel alloy, Wear. 162 (1993) 202–210.

[6] M. Abedini, H.M. Ghasemi, M.N. Ahmadabadi, Tribological behavior of NiTi alloy in martensitic and austenitic states, Mater. Des. 30 (2009) 4493–4497.

[7] W.J. Buehler, Intermetallic Compound Based Materials for Structural Applications, in: US Nav. Ord. Lab. Silver Springs, Maryland, Seventh Navy Sci. Symp. Solut. to Navy Probl. Through Adv. Technol. May, 1963: pp. 15–16.

[8]  W.J. Buehler, F.E. Wang, A summary of recent research on the Nitinol alloys and their potential application in ocean engineering, Ocean Eng. 1 (1968) 105IN7109--108IN10120.

[9]  Q. Zeng, X. Zhao, G. Dong, H. Wu, Lubrication properties of Nitinol 60 alloy used as high-speed rolling bearing and numerical simulation of flow pattern of oil-air lubrication, Trans. Nonferrous Met. Soc. China. 22 (2012) 2431–2438.

[10] Q. Zeng, G. Dong, Influence of load and sliding speed on super-low friction of nitinol 60 alloy under castor oil lubrication, Tribol. Lett. 52 (2013) 47–55.

[11] M.K. Stanford, Charpy Impact Energy and Microindentation Hardness of 60-NITINOL, NASA/TM—2012-216029 (September 2012).

[12] J. Frenzel, E.P. George, A. Dlouhy, C. Somsen, M.-X. Wagner, G. Eggeler, Influence of Ni on martensitic phase transformations in NiTi shape memory alloys, Acta Mater. 58 (2010) 3444–3458.

[13] B.C. Hornbuckle, X.Y. Xiao, R.D. Noebe, R. Martens, M.L. Weaver, G.B. Thompson, Hardening behavior and phase decomposition in very Ni-rich Nitinol alloys, Mater. Sci. Eng. A. 639 (2015) 336–344.

[14] M. Nishida, C.M. Wayman, T. Honma, Precipitation processes in near-equiatomic TiNi shape memory alloys, Metall. Trans. A. 17 (1986) 1505–1515.

[15] Q.C. Fan, Y.H. Zhang, Y.Y. Wang, M.Y. Sun, Y.T. Meng, S.K. Huang, Y.H. Wen, Influences of transformation behavior and precipitates on the deformation behavior of Ni-

rich NiTi alloys, Mater. Sci. Eng. A. 700 (2017) 269–280.

[16] G.X. Xu, L.J. Zheng, F.X. Zhang, H. Zhang, Influence of solution heat treatment on the microstructural evolution and mechanical behavior of 60NiTi, J. Alloys Compd. 775 (2019) 698–706.

[17] I. Baker, A review of the mechanical properties of B2 compounds, Mater. Sci. Eng. A. 192 (1995) 1–13.

[18] M. Karimzadeh, M.R. Aboutalebi, M.T. Salehi, S.M. Abbasi, M. Morakabati, Adjustment of aging temperature for reaching superelasticity in highly Ni-rich Ti-51.5 Ni NiTi shape memory alloy, Mater. Manuf. Process. 31 (2016) 1014–1021.

[19] Y. Zheng, F. Jiang, L. Li, H. Yang, Y. Liu, Effect of ageing treatment on the transformation behaviour of Ti–50.9 at.% Ni alloy, Acta Mater. 56 (2008) 736–745.

[20] W. Tirry, D. Schryvers, Quantitative determination of strain fields around Ni4Ti3 precipitates in NiTi, Acta Mater. 53 (2005) 1041–1049.

[21] B.C. Hornbuckle, R.D. Noebe, G.B. Thompson, Influence of Hf solute additions on the precipitation and hardenability in Ni-rich NiTi alloys, J. Alloys Compd. 640 (2015) 449–454.

[22] F. Zhang, L. Zheng, F. Wang, H. Zhang, Effects of Nb additions on the precipitate morphology and hardening behavior of Ni-rich Ni 55 Ti 45 alloys, J. Alloys Compd. 735 (2018) 2453–2461.

[23] Y. Chen, H.C. Jiang, S.W. Liu, L.J. Rong, X.Q. Zhao, The effect of Mo additions to high

damping Ti–Ni–Nb shape memory alloys, Mater. Sci. Eng. A. 512 (2009) 26–31.

[24] L. Casalena, A.N. Bucsek, D.C. Pagan, G.M. Hommer, G.S. Bigelow, M. Obstalecki, R.D. Noebe, M.J. Mills, A.P. Stebner, Structure-Property Relationships of a High Strength Superelastic NiTi–1Hf Alloy, Adv. Eng. Mater. 20 (2018) 1800046.

[25] K. Khanlari, M. Ramezani, P. Kelly, P. Cao, T. Neitzert, Comparison of the reciprocating sliding wear of 58Ni39Ti-3Hf alloy and baseline 60NiTi, Wear. 408 (2018) 120–130.

[26] O. Benafan, G.S. Bigelow, A. Garg, R.D. Noebe, Viable low temperature shape memory alloys based on Ni-Ti-Hf formulations, Scr. Mater. 164 (2019) 115–120.

[27] X.D. Han, R. Wang, Z. Zhang, D.Z. Yang, A new precipitate phase in a TiNiHf high temperature shape memory alloy, Acta Mater. 46 (1998) 273–281.

[28] B. Amin-Ahmadi, J.G. Pauza, A. Shamimi, T.W. Duerig, R.D. Noebe, A.P. Stebner, Coherency strains of H-phase precipitates and their influence on functional properties of nickel-titanium-hafnium shape memory alloys, Scr. Mater. 147 (2018) 83–87.

[29] B. Amin-Ahmadi, T. Gallmeyer, J.G. Pauza, T.W. Duerig, R.D. Noebe, A.P. Stebner, Effect of a pre-aging treatment on the mechanical behaviors of Ni 50.3 Ti 49.7-x Hfx (x 9at.%) Shape memory alloys, Scr. Mater. 147 (2018) 11–15.

[30] S.H. Mills, R.D. Noebe, C. Dellacorte, B. Amin-Ahmadi, A.P. Stebner, Development of Nickel-Rich Nickel--Titanium--Hafnium Alloys for Tribological Applications, Shape Mem. Superelasticity. (2020) 1–12.

[31] S. Mills, Development of nickel-titanium-hafnium alloys for impact resistant tribology

performances, 2019.

[32] M.J. Hÿtch, J.-L. Putaux, J.-M. Pénisson, Measurement of the displacement field of dislocations to 0.03 Å by electron microscopy, Nature. 423 (2003) 270–273. https://doi.org/10.1038/nature01638.

[33] R.A. Young, The rietveld method, International union of crystallography, 1993.

[34] B.H. Toby, R.B. Von Dreele, GSAS-II: the genesis of a modern open-source all purpose crystallography software package, J. Appl. Crystallogr. 46 (2013) 544–549.

[35] J. Khalil-Allafi, W.W. Schmahl, M. Wagner, H. Sitepu, D.M. Toebbens, G. Eggeler, The influence of temperature on lattice parameters of coexisting phases in NiTi shape memory alloys—a neutron diffraction study, Mater. Sci. Eng. A. 378 (2004) 161–164.

[36] K. Otsuka, X. Ren, Recent developments in the research of shape memory alloys, Intermetallics. 7 (1999) 511–528.

[37] F. Yang, D.R. Coughlin, P.J. Phillips, L. Yang, A. Devaraj, L. Kovarik, R.D. Noebe, M.J. Mills, Structure analysis of a precipitate phase in an Ni-rich high-temperature NiTiHf shape memory alloy, Acta Mater. 61 (2013) 3335–3346.

[38] D.R. Coughlin, L. Casalena, F. Yang, R.D. Noebe, M.J. Mills, Microstructure–property relationships in a high-strength 51Ni–29Ti–20Hf shape memory alloy, J. Mater. Sci. 51 (2016) 766–778.

[39] T. Tadaki, Y. Nakata, K. Shimizu, K. Otsuka, Crystal structure, composition and morphology of a precipitate in an aged Ti-51 at% Ni shape memory alloy, Trans. Japan

Inst. Met. 27 (1986) 731–740.

[40] R. Bonnet, E. Cousineau, D.H. Warrington, Determination of near-coincident cells for hexagonal crystals. Related DSC lattices, Acta Crystallogr. Sect. A Cryst. Physics, Diffraction, Theor. Gen. Crystallogr. 37 (1981) 184–189.

[41] M. Shamsuzzoha, R. Rahman, A New Geometrical Method for Constructing Coincident Site Lattices for Cubic Crystals, Int. J. Eng. Res. Dev. 3 (2012) 9–15.

[42] J. Khalil-Allafi, A. Dlouhy, G. Eggeler, Ni4Ti3-precipitation during aging of NiTi shape memory alloys and its influence on martensitic phase transformations, Acta Mater. 50 (2002) 4255–4274.

[43] O. Benafan, A. Garg, R.D. Noebe, H.D. Skorpenske, K. An, N. Schell, Deformation characteristics of the intermetallic alloy 60NiTi, Intermetallics. 82 (2017) 40–52.

[44] S. Pourbabak, X. Wang, D. Van Dyck, B. Verlinden, D. Schryvers, Ni cluster formation in low temperature annealed Ni 50.6 Ti 49.4, Funct. Mater. Lett. 10 (2017) 1740005.

[45] M. Karlík, P. Haušild, M. Klementová, P. Novák, P. Beran, L. Perrière, J. Kopeček, TEM phase analysis of NiTi shape memory alloy prepared by self-propagating high-temperature synthesis, Adv. Mater. Process. Technol. 3 (2017) 58–69.

[46] X. Han, W. Zou, R. Wang, S. Jin, Z. Zhang, T. Li, D. Yang, Microstructure of TiNi shape-memory alloy synthesized by explosive shock-wave compression of Ti–Ni powder mixture, J. Mater. Sci. 32 (1997) 4723–4729.

[47] D.B. Williams, C.B. Carter, The transmission electron microscope, in: Transm. Electron

Microsc., Springer, 1996: pp. 3–17.

[48] D.R. Coughlin, P.J. Phillips, G.S. Bigelow, A. Garg, R.D. Noebe, M.J. Mills, Characterization of the microstructure and mechanical properties of a 50.3 Ni–29.7 Ti–20Hf shape memory alloy, Scr. Mater. 67 (2012) 112–115.

[49] H.E. Karaca, S.M. Saghaian, G. Ded, H. Tobe, B. Basaran, H.J. Maier, R.D. Noebe, Y.I. Chumlyakov, Effects of nanoprecipitation on the shape memory and material properties of an Ni-rich NiTiHf high temperature shape memory alloy, Acta Mater. 61 (2013) 7422–7431.

[50] S.M. Saghaian, H.E. Karaca, H. Tobe, A.S. Turabi, S. Saedi, S.E. Saghaian, Y.I. Chumlyakov, R.D. Noebe, High strength NiTiHf shape memory alloys with tailorable properties, Acta Mater. 134 (2017) 211–220.

[51] T. Saburi, S. Nenno, T. Fukuda, Crystal structure and morphology of the metastable X phase in shape memory Ti-Ni alloys, J. Less Common Met. 125 (1986) 157–166.

[52] D.Y. Li, L.Q. Chen, Selective variant growth of coherent Ti11Ni14 precipitate in a TiNi alloy under applied stresses, Acta Mater. 45 (1997) 471–479.

[53] W. Tirry, D. Schryvers, High resolution TEM study of Ni4Ti3 precipitates in austenitic Ni51Ti49, Mater. Sci. Eng. A. 378 (2004) 157–160.

[54] M. Nishida, C.M. Wayman, T. Honma, Electron microscopy studies of the all-around shape memory effect in a Ti-51.0 at.% Ni alloy, Scr. Metall. 18 (1984) 1389–1394.

[55] J. Khalil-Allafi, G. Eggeler, A. Dlouhy, W.W. Schmahl, C. Somsen, On the influence of

heterogeneous precipitation on martensitic transformations in a Ni-rich NiTi shape memory alloy, Mater. Sci. Eng. A. 378 (2004) 148–151.

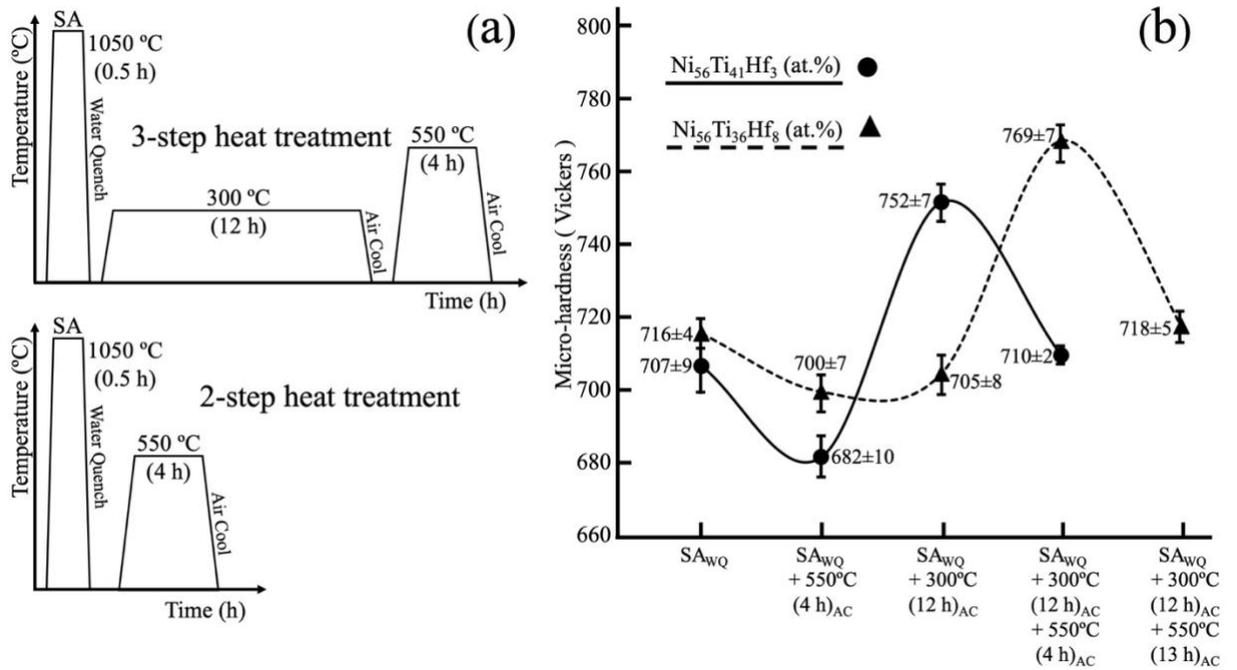

**Fig. 1.** (a) Schematic illustration of the sequence of heat treatment steps applied to the Ni$_{56}$Ti$_{36}$Hf$_8$ and Ni$_{56}$Ti$_{41}$Hf$_3$ alloys and (b) corresponding average hardness values. Solution annealing step at 1050 °C for 0.5 h is abbreviated as "SA". Note, the lines connecting the data points are for visualization purposes.

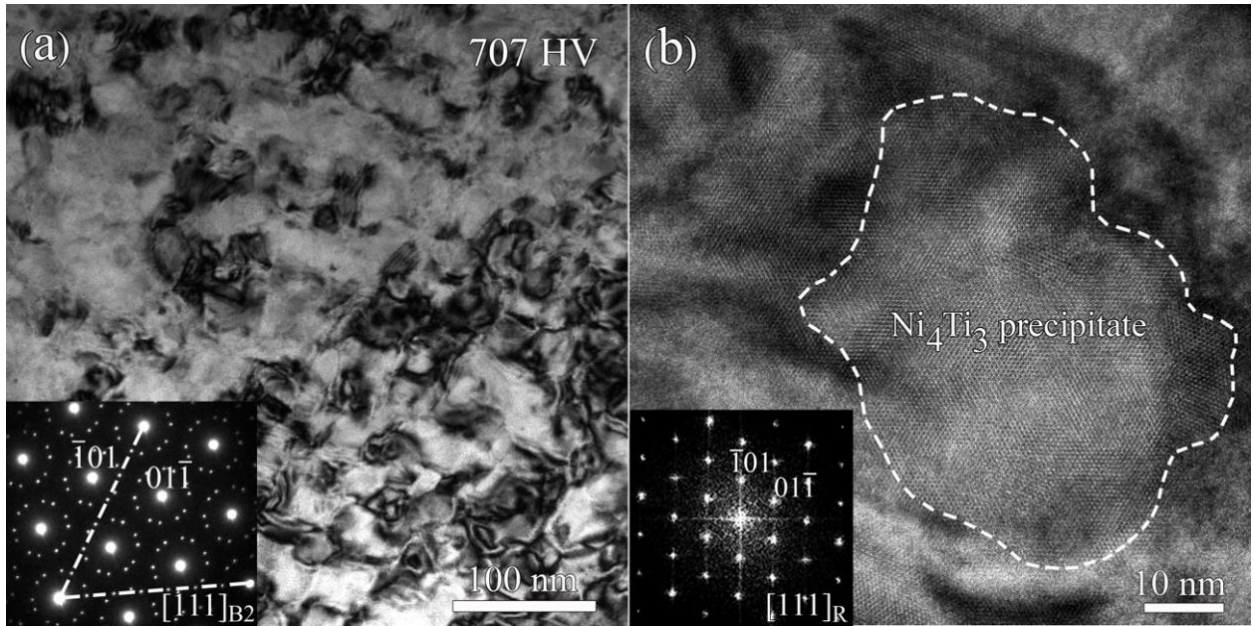

**Fig. 2.** (a) Conventional BF-TEM micrograph of Ni$_{56}$Ti$_{41}$Hf$_3$ after solution annealing at 1050 °C for 0.5 h followed by a water quench. The corresponding SAED pattern in the bottom left inset, taken along [111]$_{B2}$ zone axis, shows the super reflections originated from two variants of Ni$_4$Ti$_3$ precipitates. (b) HRTEM micrograph taken along ([111]$_R$/ /[111]$_{B2}$) showing a monolithic rhombohedral (R) Ni$_4$Ti$_3$ precipitate within the B2 matrix.

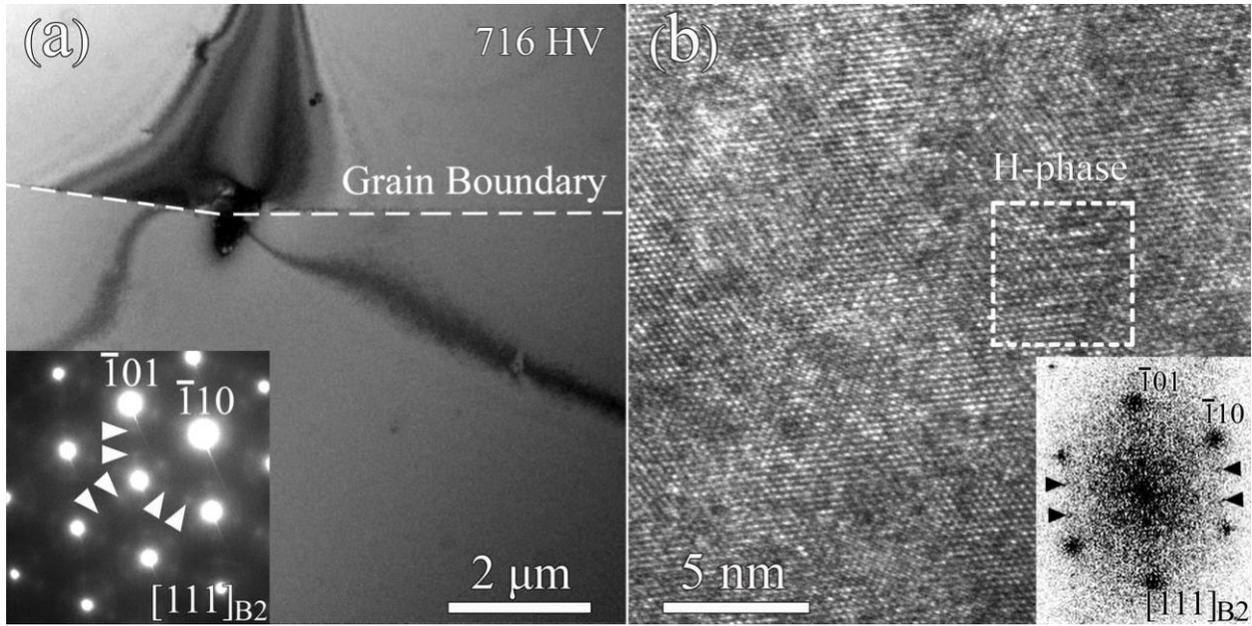

**Fig. 3.** (a) BF-TEM micrograph of Ni$_{56}$Ti$_{36}$Hf$_{8}$ after solution annealing at 1050 °C for 0.5 h followed by a water quench. The corresponding SAED pattern in the bottom left inset shows weak super reflections along 1/3<011> (indicated by arrowheads), which originate from 3 different variants of H-phase precipitates. (b) HRTEM micrograph taken along [111]$_{B2}$ showing a small H-phase precipitate. The super reflections due to the H-phase precipitate are indicated by arrowheads in the FFT pattern.

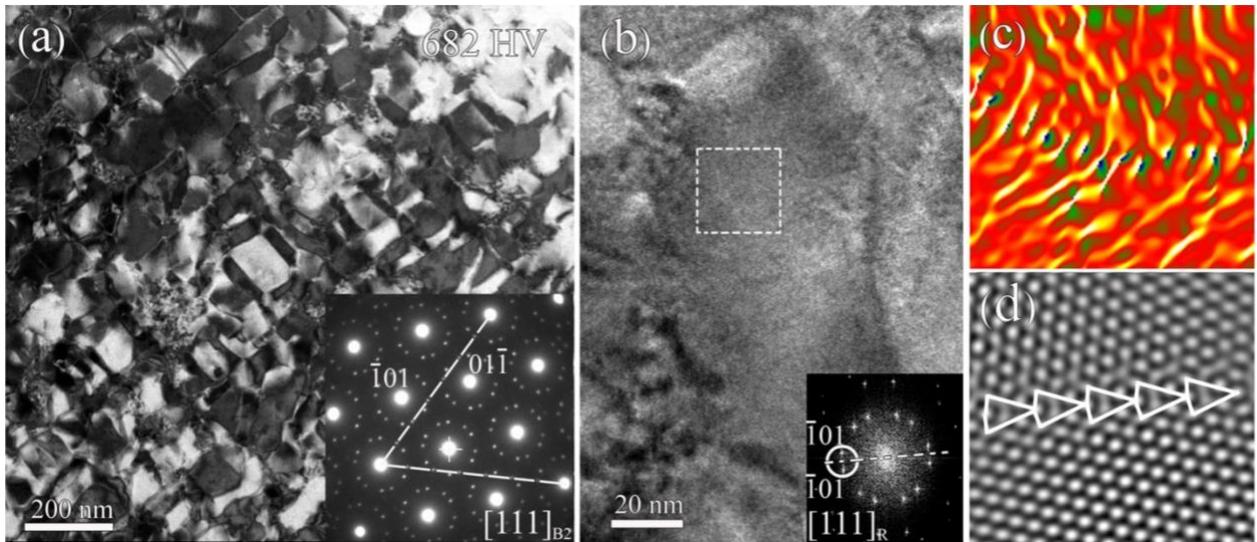

**Fig. 4.** (a) BF-TEM micrograph and corresponding SAED pattern (inset) of $Ni_{56}Ti_{41}Hf_3$ after solution annealing and aging at 550 ºC for 4 h showing the blocky morphology of $Ni_4Ti_3$ precipitates within the B2 matrix. (b) HRTEM micrograph taken along $[111]_R$ zone axis showing coalesced $Ni_4Ti_3$ precipitates involving two variants. Corresponding FFT pattern (inset) taken from the region indicated by the dashed box shows the twin relation (white dashed line) between two variants along the boundary. (c) Local **g**-map using $\mathbf{g} = (\bar{1}01)_R$ and $\mathbf{g} = (\bar{1}01)_R$ (indicated by white circle on the FFT pattern) was taken from the region indicated by the dashed box in (b). The misfit dislocations along the boundary are shown as hot spots. (d) Enlarged IFFT filtered image of the interface showing individual structural units of the $\Sigma = 7$ CSL type interface.

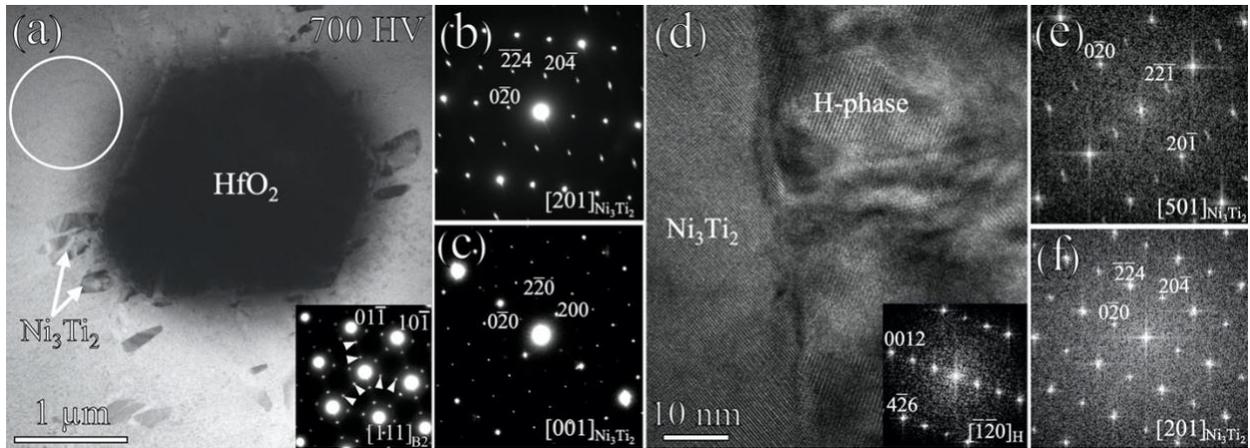

**Fig. 5.** (a) BF-TEM micrograph of Ni56Ti36Hf8 after being solution annealed and aged at 550 ºC for 4 h. The corresponding SAED pattern (inset) was taken along the [111]$_{B2}$ zone from the region indicated by the white circle and contains super reflections (white arrowheads) originating from H-phase precipitates. SAED patterns taken from large Ni3Ti2 precipitates surrounding HfO2 along (b) [201]$_{Ni3Ti2}$ zone and (c) [001]$_{Ni3Ti2}$ zone confirm orthorhombic Ni3Ti2 structure. (d) HRTEM micrograph taken along [111]$_{B2}$ showing Ni3Ti2 phase (left) and H-phase precipitates within B2 matrix (right) which was identified from the FFT (inset) taken on the $[\overline{1}\overline{2}0]_H$ zone of face centered orthorombic. (c) FFT taken from Ni3Ti2 precipitate along the [501]$_{Ni3Ti2}$ zone confirms the orthorhombic structure . An additional FFT taken from an Ni3Ti2 precipitate along [201]$_{Ni3Ti2}$ zone from a different HRTEM micrograph (not shown) further confirms orthorhombic structure of the large Ni3Ti2 phase.

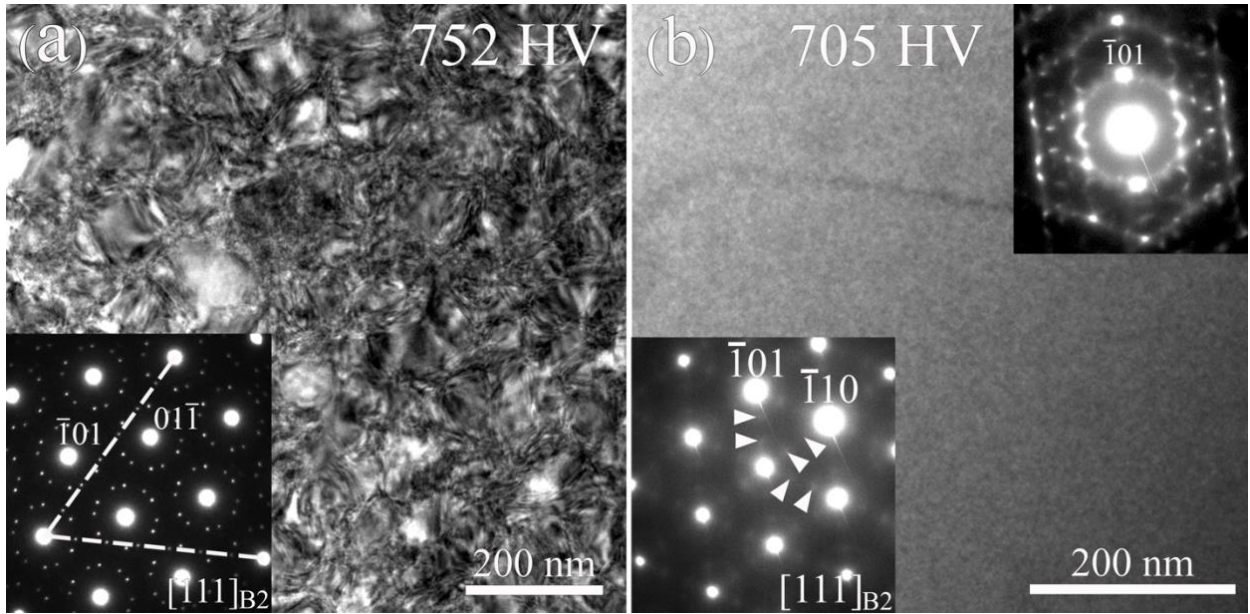

**Fig. 6.** Conventional BF-TEM micrographs and corresponding SAED patterns after solution annealing and preaging at 300 ºC for 12 h. (a) the $Ni_{56}Ti_{41}Hf_3$ alloy containing multi-variant equiaxed $Ni_4Ti_3$ precipitates and (b) the nearly featureless $Ni_{56}Ti_{36}Hf_8$ alloy exhibiting faint super reflections along 1/3<110> from a low density of fine H-phase precipitates. The top right inset shows an SAED pattern tilted slightly off the $[111]_{B2}$ zone, revealing diffuse reflections likely due to Hf and Ni clustering.

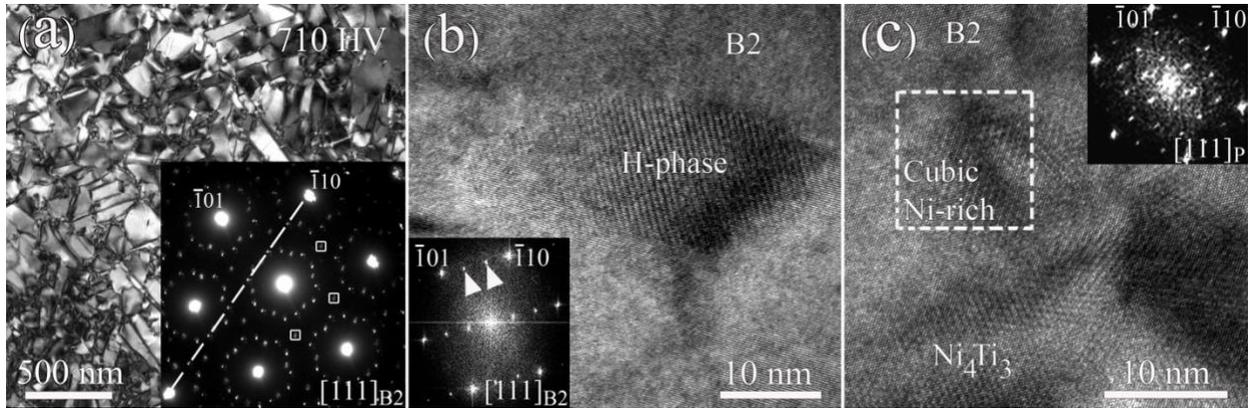

**Fig. 7.** (a) Conventional BF-TEM micrograph and corresponding SAED pattern (inset) from the Ni$_{56}$Ti$_{41}$Hf$_3$ alloy after solution annealing, preaging at 300 °C for 12 h and aging at 550 °C for 4 h. A set of super reflections from the Ni$_4$Ti$_3$ precipitates are indicated by a white dashed line, super reflections along 1/3<110> common between H-phase and cubic Ni-rich precipitates and super-reflections unique to cubic Ni-rich precipitates (indicated by white boxes) are also observed. (b) HRTEM micrograph taken along [111]$_{B2}$ zone axis shows a monolithic H-phase precipitate surrounded by B2 matrix. Super reflections in the FFT pattern originating from the H-phase precipitate are indicated by arrowheads. (c) HRTEM micrograph taken along [111]$_{B2}$ zone and corresponding FFT pattern confirm the existence of a cubic Ni-rich precipitate surrounded by B2 matrix. Cubic Ni-rich precipitate is identified as "P" in the SAED and FFT patterns throughout this study. A larger Ni$_4$Ti$_3$ precipitate is observed at the bottom of the micrograph.

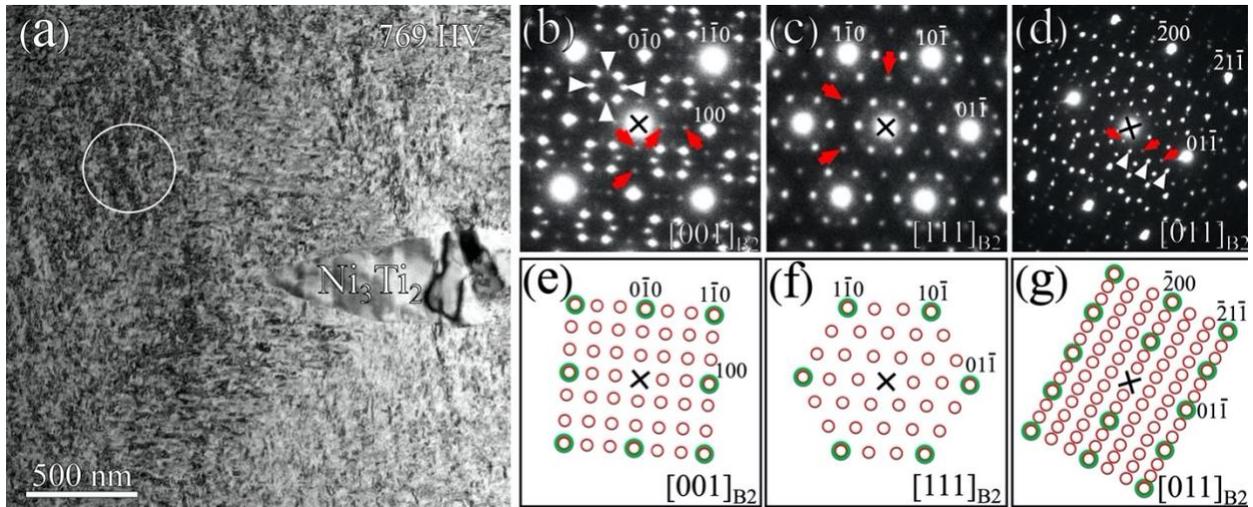

**Fig. 8.** (a) A conventional BF-TEM micrograph of Ni$_{56}$Ti$_{36}$Hf$_8$, solution annealed at 1050 ºC for 0.5 h, pre-aged at 300 ºC for 12 h and aged at 550 ºC for 4 h, shows a mottled microstructure consisting of a mixture of nano-scale precipitates in a B2 matrix. SAED patterns taken along (b) [001]$_{B2}$, (c) [111]$_{B2}$ and (d) [001]$_{B2}$ zones show the coexistence of B2-phase (selected reflections are indexed), H-phase (unique superlattice reflections are indicated with white arrows), and cubic Ni-rich phase (unique superlattice reflections indicated with red arrows), the remaining superlattice reflections are common between H-phase and cubic Ni-rich phase. (e)–(f) Schematics of diffraction spot arrangements from cubic Ni-rich (red) in correspondence with B2 matrix spots (green) taken along [001]$_{B2}$, [111]$_{B2}$ and [011]$_{B2}$ zones.

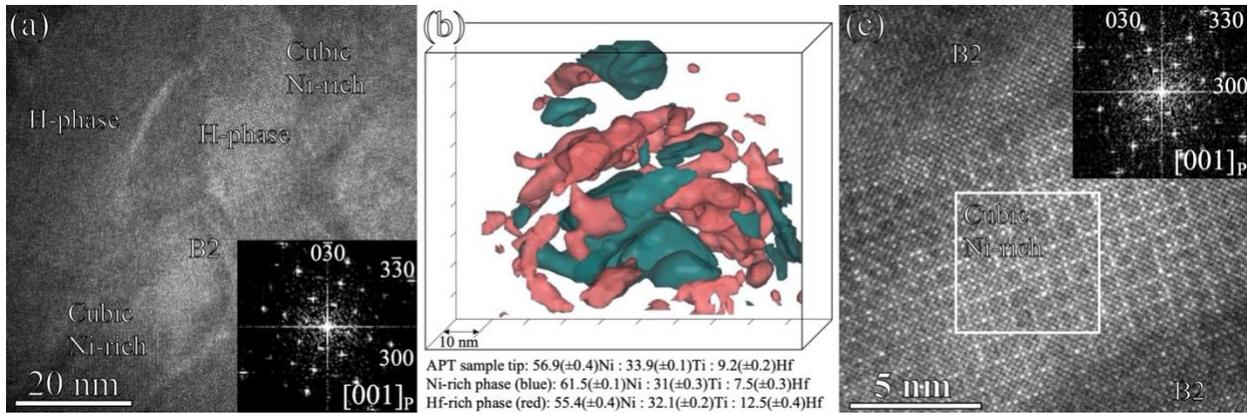

**Fig. 9.** (a) HR-TEM micrograph taken along [001]$_{B2}$ zone axis in Ni$_{56}$Ti$_{36}$Hf$_8$ alloy after a three-step heat treat (solution annealed, preaged and aged at 550 °C for 4 h) reveals dense nanoprecipitation of H-phase and cubic Ni-rich precipitates (identified by FFT) with narrow B2 channels. (b) APT reconstruction tip composed of mottled 3-phase microstructure 60 at.% Ni iso-composition surface showing the morphology and distribution of 3-phase microstructure containing cubic Ni-rich phase (dark blue), Hf-rich phase (light red). Average composition of cubic Ni-rich precipitates and H-phase precipitates are listed along with the bulk APT sample tip composition. (c) Atomic resolution HAADF-STEM image along [001]$_{B2}$ of new cubic Ni-rich phase within the B2 matrix. Corresponding FFT (inset) taken from the region indicated by white square is a pattern characteristic of cubic Ni-rich phase. The repeating grid-like arrangement of higher Z-contrast atoms provides additional insight to the precipitate structure and atomic arrangement.

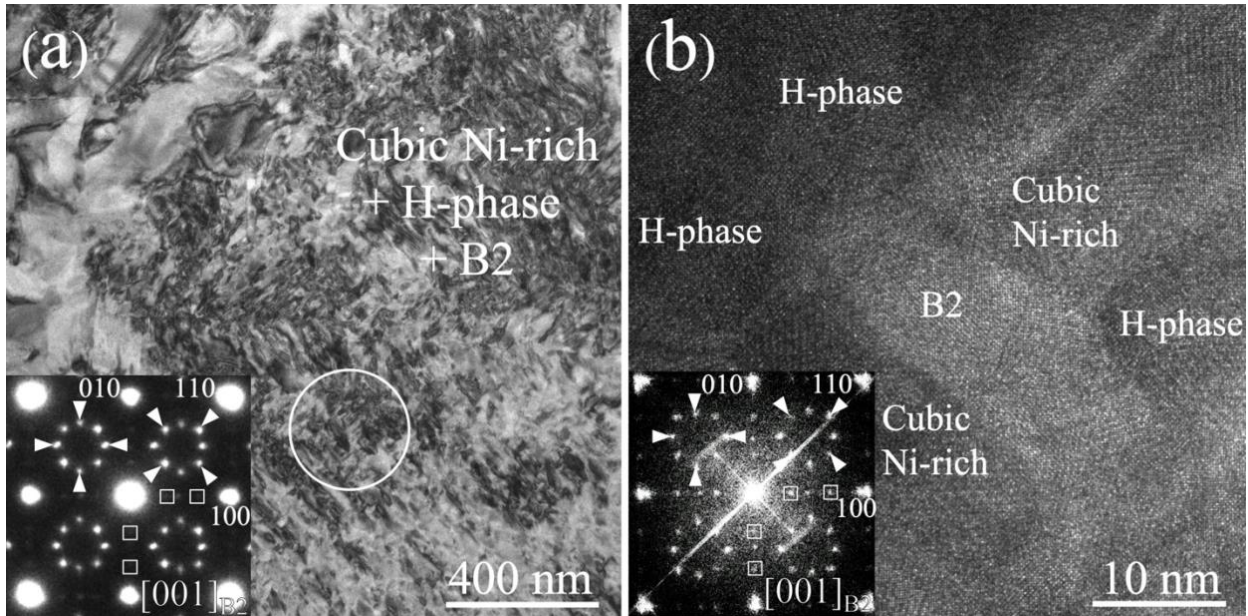

**Fig. 10.** (a) BF-TEM micrograph of the $Ni_{56}Ti_{36}Hf_8$ alloy after a three-step heat treat (solution annealed, preaged and aged at 550 ºC for 13 h) showing a mottled microstructure. The SAED pattern along $[001]_{B2}$ zone axis (inset) taken from the region indicated by white circle shows the existence of H-phase (white arrows) and cubic Ni-rich precipitates (white squares). (b) HR-TEM micrograph taken along $[001]_{B2}$ zone axis and corresponding FFT (inset) reveals dense nanoprecipitation of H-phase and small cubic Ni-rich precipitates with narrow channels of B2 matrix.

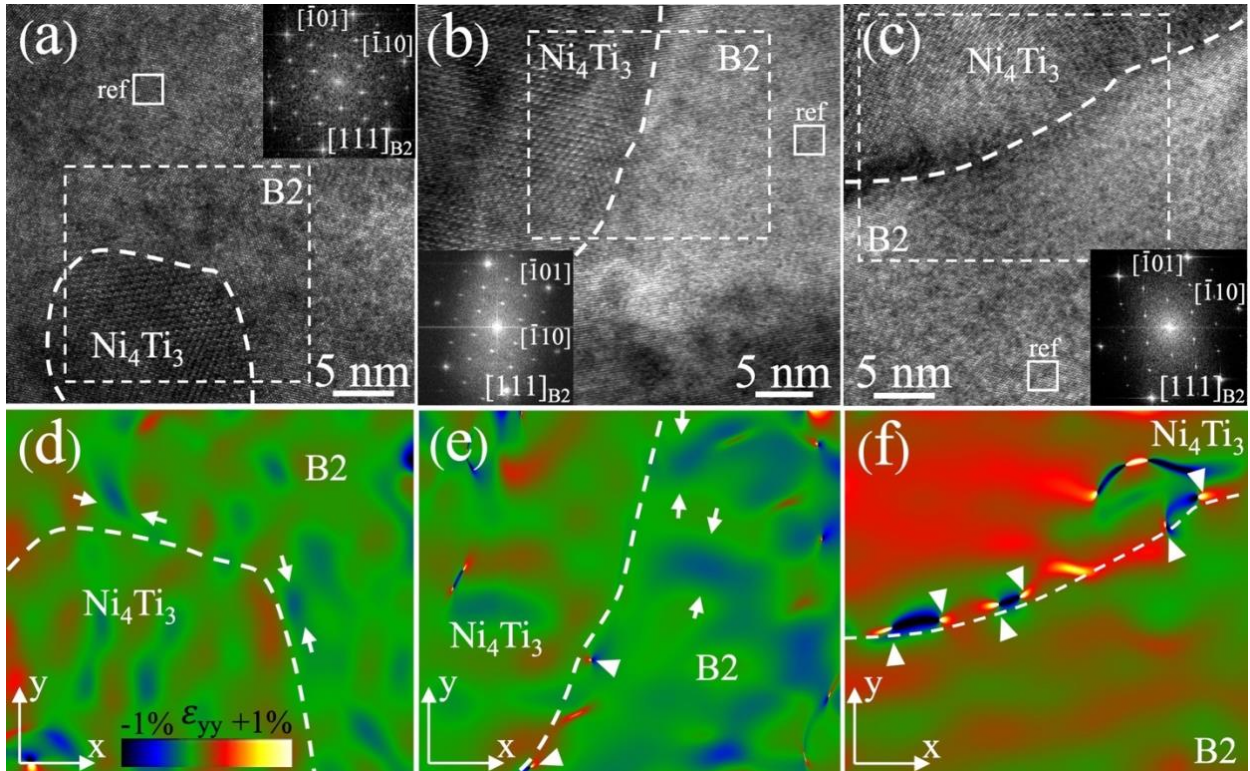

**Fig. 11.** HR-TEM micrographs taken along the [111]$_{B2}$ zone axis showing Ni$_4$Ti$_3$ / matrix interfaces in the Ni$_{56}$Ti$_{41}$Hf$_3$ alloy after: (a) SA$_{WQ}$, (b) SA$_{WQ}$ + 300°C (12 h) (c) SA$_{WQ}$ + 300°C (12 h) + 550°C (4 h). (d)–(f) Corresponding GPA maps of the boxed regions in (a)-(c), respectively, showing the magnitude of the $\varepsilon_{yy}$ strain component. White arrowheads indicate misfit dislocations along the interface.

**Table 1.** Precipitate properties for the $Ni_{56}Ti_{41}Hf_3$ and $Ni_{56}Ti_{36}Hf_8$ alloys after application of the various heat treatment steps (length (L), width (W), interparticle distance (D) and area fraction (A %) are included for each precipitate phase observed). All precipitate dimensions are reported in nanometers (nm) and the hardness values are for the bulk alloy.

| Composition / HT | $Ni_4Ti_3$ (L) | $Ni_4Ti_3$ (W) | $Ni_4Ti_3$ (D) | $Ni_4Ti_3$ (A%) | H-phase (L) | H-phase (W) | H-phase (D) | H-phase (A%) | Cubic Ni-rich (L) | Cubic Ni-rich (W) | Cubic Ni-rich (D) | Cubic Ni-rich (A %) | Hardness (HV) |
|---|---|---|---|---|---|---|---|---|---|---|---|---|---|
| $Ni_{56}Ti_{41}Hf_3$ SA | 51 ± 18 | 51 ± 18 | 26 ± 9 | 54 | ** | ** | ** | ** | ** | ** | ** | ** | 707 ± 9 |
| $Ni_{56}Ti_{41}Hf_3$ SA + 550(4 h) | 109 ± 33 | 87 ± 19 | 92 ± 24 | 61 | ** | ** | ** | ** | ** | ** | ** | ** | 682 ± 10 |
| $Ni_{56}Ti_{41}Hf_3$ SA + 300(12 h) | 81 ± 15 | 81 ± 15 | 34 ± 19 | 71 | ** | ** | ** | ** | ** | ** | ** | ** | 752 ± 7 |
| $Ni_{56}Ti_{41}Hf_3$ SA + 300(12 h) + 550(4 h) | 138 ± 41 | 94 ± 23 | 105 ± 33 | 63 | 28 ± 4 | 13 ± 2 | 104 ± 13 | 3 | 3-7 | 3-7 | ** | < 1 | 710 ± 2 |
| $Ni_{56}Ti_{36}Hf_8$ / SA | ** | ** | ** | ** | 2 - 5 | 2 - 5 | ** | < 1 | ** | ** | ** | ** | 716 ± 4 |
| $Ni_{56}Ti_{36}Hf_8$ SA + 550(4 h) | ** | ** | ** | ** | 21 ± 6 | 8 ± 2 | 17 ± 6 | 49 | ** | ** | ** | ** | 700 ± 7 |
| $Ni_{56}Ti_{36}Hf_8$ SA + 300(12 h) | ** | ** | ** | ** | 2 - 5 | 2 - 5 | ** | < 1 | ** | ** | ** | < 1 | 705 ± 8 |
| $Ni_{56}Ti_{36}Hf_8$ SA + 300(12 h) + 550(4 h) | ** | ** | ** | ** | 23 ± 5 | 12 ± 3 | 7 ± 2 | 54 | 18 ± 4 | 16 ± 5 | 11 ± 7 | 33 | 769 ± 7 |
| $Ni_{56}Ti_{36}Hf_8$ SA + 300(12 h) + 550(13 h) | ** | ** | ** | ** | 40 ± 8 | 21 ± 6 | 14 ± 4 | 61 | 20 ± 4 | 17 ± 3 | 22 ± 7 | 19 | 718 ± 5 |